# Diminished Diversity-of-Thought in a Standard Large Language Model

Peter S. Park[1, *], Philipp Schoenegger[2, *], Chongyang Zhu[3]

[1] Department of Physics, MIT, 70 Vassar Street, Cambridge, MA, USA (dr_park@mit.edu)

[2] Department of Management, London School of Economics, Marshall Building, 44 Lincoln's Inn Fields, London, England, UK (contact.schoenegger@gmail.com)

[3] CVS Health, 1 CVS Dr., Woonsocket, RI, USA (cyzhu95@gmail.com)

[*] Co-first author

**Abstract:** We test whether Large Language Models (LLMs) can be used to simulate human participants in social-science studies. To do this, we run replications of 14 studies from the Many Labs 2 replication project with OpenAI's *text-davinci-003* model, colloquially known as GPT3.5. Based on our pre-registered analyses, we find that among the eight studies we could analyse, our GPT sample replicated 37.5% of the original results and 37.5% of the Many Labs 2 results. However, we were unable to analyse the remaining six studies due to an unexpected phenomenon we call the *"correct answer" effect*. Different runs of GPT3.5 answered nuanced questions probing political orientation, economic preference, judgement, and moral philosophy with zero or near-zero variation in responses: with the supposedly "correct answer." In one exploratory follow-up study, we found that a "correct answer" was robust to changing the demographic details that precede the prompt. In another, we found that most but not all "correct answers" were robust to changing the order of answer choices. One of our most striking findings occurred in our replication of the Moral Foundations Theory survey results, where we found GPT3.5 identifying as a political conservative in 99.6% of the cases, and as a liberal in 99.3% of the cases in the reverse-order condition. However, both self-reported 'GPT conservatives' and 'GPT liberals' showed right-leaning moral foundations. Our results cast doubts on the validity of using LLMs as a general replacement for human participants in the social sciences. Our results also raise concerns that a hypothetical AI-led future may be subject to a diminished diversity-of-thought.

# 1. Introduction

The field of Natural Language Processing (NLP) has witnessed rapid advances. This trend is most recently exemplified by Large Language Models (LLMs). When trained on large corpora of internet- and book-based text data to predict the next sequence of words given an input, LLMs have demonstrated the ability to generate sophisticated responses to a wide range of prompts. OpenAI's GPT-3 family of models (Brown et al., 2020), its successor GPT-4 (OpenAI, 2023b), and the models' chatbot version ChatGPT (OpenAI, 2023c) have received significant attention, in particular due to the models' capabilities in a wide variety of tasks that were previously thought to require human intelligence (Metz, 2020). To illustrate, GPT-4 has excelled on versions of difficult standardized tests originally meant for humans (OpenAI, 2023b), although it is sometimes unclear whether its answers to these tests were memorized from the training data. GPT-4 has even shown an arguably human-rivalling ability to solve potentially novel tasks in vision, mathematics, coding, medicine, and law (Bubeck et al., 2023). Companies are already using OpenAI's models to automate economically valuable services, such as the presentation of information via search-engine chatbots (Roose, 2023), via AI personal assistants (Warren & Lawler, 2023), and even via the writing of media content (Edwards, 2023).

OpenAI's mission is to create "highly autonomous systems that outperform humans at most economically valuable work" (OpenAI, 2020). Regardless of whether or when this mission will be achieved, by either OpenAI or its competitors, people are prone to treating even current LLMs as if they possess human-like qualities: an anthropomorphisation that is not always rigorously justified or investigated (Salles et al., 2020). Because of the potentially sweeping societal changes that advanced AI may bring with it and the anthropomorphisation of current LLM models, the rigorous study of these models, their applications, and limitations are especially critical.

One way that LLMs have been studied before in the social sciences is by studying them with the methods of psychology as if they were human participants, and potentially even as "surrogates" (Grossman et al., 2023, p. 1108) that directly supplant human participants. Much of this previous work has implicitly or explicitly assumed that concepts from the psychological sciences and experimental methods originally meant for humans can be applied straightforwardly to LLMs: to elicit supposedly parallel mechanisms of human and LLM cognition, to psychologically categorise LLMs as if they were humans, and even to simulate human behavioural data (Dillion et al., 2023). To illustrate, Binz and Schulz (2023) conducted vignette-based survey experiments on GPT-3 and concluded from their data that the LLM showed signs of model-based reinforcement learning and of behavioural similarities to humans. Miotto et al. (2022) investigated GPT-3's personality characteristics, values, and self-reported demographic properties. Similarly, Li et al. (2022) investigated the personality of GPT-3 using the Short Dark Triad scale of narcissism, psychopathy, and Machiavellianism (Jones & Paulhus, 2014); and the Big Five inventory of openness to experience, conscientiousness, extraversion, agreeableness, and neuroticism (John & Srivastava, 1999). Horton (2023) examined GPT-3 in the context of behavioural-economics experiments and concluded that its behaviour was qualitatively similar to that of human participants. Shihadeh et al. (2022) measured the presence of the brilliance bias—the bias that brilliance is seen as a male trait—in GPT-3. Finally, Argyle et al. (2022) and Aher et al. (2022) each conducted similar experiments in which different types of participants were simulated via GPT-3 and proposed that the model may be used to indirectly collect data on the behavioural aspects of various human subjects. These approaches are so far characterised by a significant heterogeneity of methods and have produced mixed results in terms of LLMs' ability to supplant human subjects.

In this paper, we conduct a multifaceted investigation of whether psychology studies originally designed for human participants can in fact be straightforwardly applied to LLMs, and whether LLMs can replace human participants in these studies. Specifically, we study OpenAI's *text-davinci-003* model (OpenAI, 2023d), a variant of GPT-3 colloquially known as GPT3.5, with a large set of psychology studies originally replicated by the Many Labs 2 project (Klein et al., 2018), a large-scale replication project in psychology. We replicate as many studies from this set as is feasible in the current monomodal context of GPT3.5, and analyse which effects successfully replicate to give us a direct and representative measure of how widely LLMs may or may not be applicable in supplanting human participants. Our replications are aided by the fact that, unlike human subjects, GPT3.5 allows for well-controlled experiments that are highly powered and unlikely to suffer from a variety of sampling, attention, and other design issues that human studies must grapple with. This is because large samples can be collected quickly and inexpensively, without sampling biases—such as non-response bias and exclusion bias—that in practice inevitably consign human samples to be insufficiently representative of the sheer diversity of human psychologies around the world (Henrich et al., 2010; Majid, 2023; Schimmelpfennig et al., 2023). Analysing the ways in which different runs of GPT3.5 answer the studies' survey questions—and how they are similar to or different from human responses—can help rigorously contribute to a broader and more interdisciplinary understanding of the AI model, its applications, and its respective limitations.

## 2. Methods

Full details of the methods can be found in the Supplementary Information. We pre-registered this study on the Open Science Framework (Park et al., 2023). For our study, we drew on the set of studies used in Many Labs 2 (Klein et al., 2018) and their respective analysis plans. The total number of potential studies that we could analyse was 28. We excluded a total of 14 studies prior to data collection as their designs included pictures, compared national samples, relied on handwriting or font changes, or used an otherwise inapplicable component that was not transferable to GPT3.5's monomodal context. We then presented GPT3.5 with these remaining surveys, with each run representing a new call to the model's API; see Figure 1 for a sample input and output. Then, we converted the survey responses of GPT3.5 runs from .txt to .csv for statistical analysis and removed all entries that responded to questions with characters that were not among the possible response categories. For example, the responses could have had characters like '?' or '/' instead of the expected outputs that we could use as unambiguous survey responses. Next, we analysed the data with the respective analysis plan that was based on that of Many Labs 2 (differences and exceptions are noted in the Supplementary Information).

Overall, we collected data for a total of 14 studies for our main pre-registered analyses, each of which consisted of about 1,000 different runs of the default temperature setting of GPT3.5, which represents the central source of variation in responses. The (softmax) temperature parameter measures the degree to which the model's outputs are predetermined. Specifically, the model's probability value of predicting the specific token (unit of text) $t_i \in \{t_1, ..., t_N\}$ to be the next token is given by a certain function of the logit value $z_i$ corresponding to the token, defined by $\mu(z_i) = \frac{e^{z_i/T}}{\sum_{k=1}^{N} e^{z_k/T}}$. This comprises a probability distribution that approaches a one-point distribution on the most probable token as $T \to 0$, and approaches an equidistribution across all tokens as $T \to \infty$. The effect of the temperature parameter on a hypothetical probability distribution for the next

**Figure 1**. Sample input and output from our GPT3.5 replication of the study of Rottenstreich and Hsee (2001). As a prompt-engineering technique, we have put—before the survey—instructions on how to format its output, and—after the survey—a "CHECKLIST FOR SURVEY ANSWERS" section to remind GPT3.5 of these instructions.

```
Input
Your task is to answer a survey question. Given the survey question, return
the [blank] replaced with your answer.
The [blank] should be replaced with only a single capitalized alphabet
letter, and nothing more.
For example, if you choose choice A, you should replace '[blank]' with 'A'
and end the line immediately afterwards.
Do your task step by step and only then return the completed survey
question.

SURVEY QUESTION:

Imagine that you have the opportunity to either meet and kiss your
favorite movie star or receive $50 in cash.
Which would you prefer?
A: Meet and kiss my favorite movie star.
B: Receive $50 in cash.
Question 1 (Multiple choice):
[blank]

CHECKLIST FOR SURVEY ANSWER:

Did you replace [blank] with your response?
Is your response exactly one capitalized letter long (A or B) with no further
explanations or symbols?

SURVEY ANSWER:
Question 1 (Multiple choice):

Output
A
```

token is illustrated in Figure 2. In accordance with likely societal use, we set the temperature parameter to the default intermediate value of 1.0 (OpenAI, 2023a).

Historically, GPT3.5's default temperature setting of 1.0 has generally been thought to output answers that are not predetermined. This can be seen from OpenAI's instruction regarding the temperature parameter: "higher values [of temperature] like 0.8 will make the output more random, while lower values like 0.2 will make it more focused and deterministic" (OpenAI, 2023a). Thus, at the time of pre-registration, we had not considered the possibility that all or nearly all ~1,000 runs of the default temperature setting of GPT3.5 might answer one of our survey questions in a predetermined way. If this were to occur, the zero or near-zero variation in this central variable would make the corresponding study's statistic—the one we had planned to analyse—unsuitable, and perhaps even unconstructable as a well-defined statistic. As such, we also conducted a number of exploratory follow-up studies to further probe our results. More details on these studies' methods are also available in the Supplementary Information.

## 3. Results

We find that surveyed runs of GPT3.5 provided responses that were in some ways comparable to those given by the corresponding human subjects. To illustrate, in the survey of Kay et al. (2014) on whether structure promotes goal pursuit, different runs of GPT3.5 gave human-like answers when asked about their long-term goal. These answers ranged from becoming fluent in Spanish, to becoming a full-time freelance software developer, to achieving financial freedom. Different runs of GPT3.5 also responded to reading-comprehension questions in the survey of Kay et al. (2023) with accurate, well-written, and grammatically correct answers, such as "Stars can turn into neutron stars, white dwarfs, or brown dwarfs" and "Light from stars takes over 100 years to reach us because of the vast distances between us and the stars." These answers were consistently on-topic, in contrast to some human subjects' tendency to sometimes write completely off-topic answers: to illustrate, consider the human answer "Y'all need a panda tail to go to bed and go get food or drinks sugar or drinks and then I eat a chicken nuggets" to the survey question "When you visit a new city, what kinds of activities do you like to do?" (Kennedy, 2020).

**Figure 2**. Effect of softmax temperature on next-token prediction probabilities. The left column's figure illustrates the original logits (with hypothetical values L=1.00, 1.25, 1.50, 1.75, 2.00) corresponding to five tokens comprising a hypothetical model. The right column's figures show the probabilities corresponding to the tokens' logits after applying the softmax function for four different temperature values (T=0.2, 0.8, 1.0, and 2.0). The figure shows how varying the temperature parameter affects the degree of predeterminedness (lower temperatures) or randomness (higher temperatures) in the model's next-token prediction.

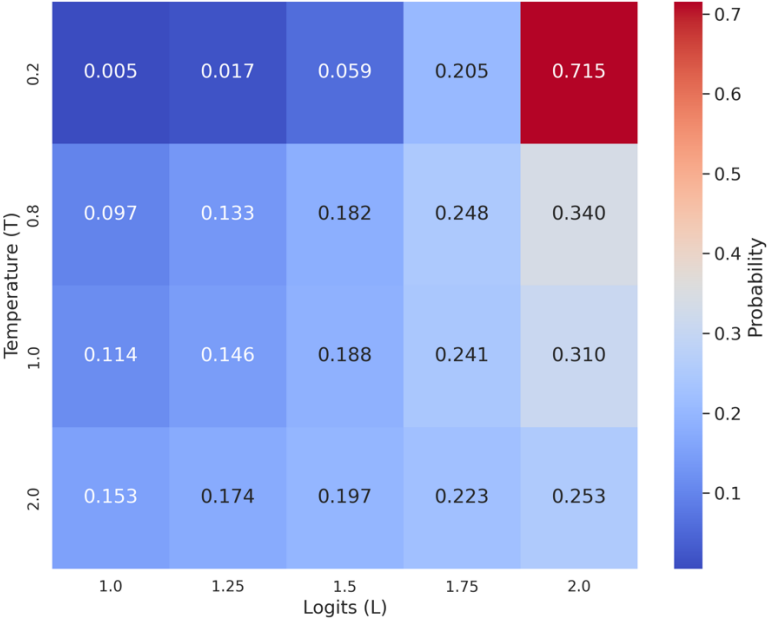

Based on our pre-registered analyses, among the eight studies we could analyse, our GPT sample replicated 37.5% of the original effects and 37.5% of the Many Labs 2 effects. Both percentages were lower than the Many Labs 2 project's 50% replication rate for the original versions of this subset of eight studies. There was substantial heterogeneity in whether our GPT sample replicated the study's original finding and whether it replicated the corresponding finding of the Many Labs 2 project. For the study of Ross et al. (1977) testing the false-consensus effect on the traffic ticket scenario and the study of Hsee (1998) on the less-is-better effect, our GPT sample successfully replicated both the original result and the corresponding Many Labs 2 result. For the study of Shafir (1993) on the effect of choosing versus rejecting on relative desirability, our GPT sample successfully replicated the original result, but did not replicate the corresponding Many Labs 2 result. For the study of Kay et al. (2014), our GPT sample successfully replicated the Many Labs 2 result but did not replicate the original result. And for all other studies we could analyse, our GPT sample did not replicate either the original result or the corresponding Many Labs 2 result.

The effect sizes found by the original studies, the Many Labs 2 replications, and our GPT replications are listed in Table 1. The verbal descriptions of the effects, whether Many Labs 2 successfully replicated the original findings, and whether our GPT sample successfully replicated the original findings and the Many Labs 2 findings, can be found in Table 2.

**Table 1**. Comparison between the Cohen's d, Cohen's q, or Odds Ratio effect sizes (with 95% confidence intervals presented in brackets) for GPT3.5, Many Labs 2, and the original results. Successful replications are bolded. Six studies could not be analysed due to the "correct answer" effect. Our result for the study of Schwarz et al. (1991) is underlined because unlike both the original sample and the Many Labs 2 sample, our GPT sample answered the two questions in a negatively correlated rather than a positively correlated manner: a qualitatively different finding.

| Study | Description | Original | ML2 | GPT |
|---|---|---|---|---|
| **Structure promotes goal pursuit** (Kay et al., 2014) | Subjects read a passage in which a natural event was described as either structured or random. The effect on subjects' willingness to pursue their goal was measured. | d=0.49 [0.001, 0.973] | d=−0.02 [−0.07, 0.03] | d=0.06 [−0.06, 0.19] |
| **Moral foundations of liberals versus conservatives** (Graham et al. 2009) | Subjects first self-identified on the liberal-conservative spectrum. The effect of this on whether concerns for the in-group, authority, or purity were thought to be more relevant for moral judgement was measured. | d=−0.43 [−0.55, −0.32] | d=0.29 [0.25, 0.34] | "correct answer" effect |
| **Affect and risk** (Rottenstreich & Hsee, 2001) | Subjects were asked to choose between a kiss from a favourite movie star and $50, either with a certain outcome or with only a 1% chance of getting the outcome. | d=0.74 [< 0.001, 1.74] | d=−0.08 [−0.13, −0.03] | "correct answer" effect |
| **Consumerism undermines trust** (Bauer et al., 2012) | Subjects read a passage in which they and others were described either as "consumers" or "individuals." The effect on whether they trusted that others would conserve water was measured. | d=0.87 [0.41, 1.34] | **d=0.12 [0.07, 0.17]** | d=0.11 [-0.01, 0.23] |
| **Disgust sensitivity predicts homophobia** (Inbar et al., 2009) | Subjects read a passage about a director encouraging homosexual versus heterosexual kissing. The effect of their disgust sensitivity on whether they considered the encouragement intentional was measured. | q=0.70 [0.05, 1.36]. | d=−0.02 [−0.06, 0.03] | d=−0.58 [−0.71, −0.45] |
| **Trolley Dilemma: principle of double effect** (Hauser et al., 2007) | Subjects were asked whether they would sacrifice one life to save five lives, either by changing an out-of-control trolley's trajectory or by pushing a large man in front of a trolley. | d=2.50 [2.22, 2.86] | **d=1.35 [1.28, 1.41]** | "correct answer" effect |
| **False consensus: supermarket scenario** (Ross et al., 1977) | In a scenario at a supermarket, subjects estimated whether they and others would sign a release for a TV commercial. Their estimated probability of signing and of others' probability were compared. | d=0.79, [0.56, 1.02] | **d=1.18, [1.13, 1.23]** | "correct answer" effect |
| **False consensus: traffic-ticket scenario** (Ross et al., 1977) | In a traffic-ticket scenario, subjects estimated whether they and others would either pay the fine or go to court. Their estimated probability of paying the fine and of others' probability were compared. | d=0.80, [0.22, 1.87] | **d=0.95, [0.90, 1.00]** | d=1.27 [1.11, 1.42] |
| **Effect of framing on decision making** (Tversky & Kahneman, 1981) | In a scenario of buying both a cheap item and an expensive item, subjects answered whether they would buy at a far-away store with a fixed discount on either the cheap item or the expensive item. | OR=4.96 [2.55, 9.90] | **OR=2.06 [1.87, 2.27]** | "correct answer" effect |
| **Reluctance to tempt fate** (Risen & Gilovich, 2008) | In the role of a student in class, subjects estimated their likelihood of being called on when being told they had not prepared for class (tempting fate) versus being told they had prepared (not tempting fate). | d=0.39, [0.03, 0.75] | **d=0.18, [0.14, 0.22]** | d=−2.49 [−2.68, −2.29] |
| **Less-is-better effect** (Hsee, 1998) | Subjects estimated the degree of generosity of a less expensive gift in a more expensive category versus a more expensive gift in a less expensive category. | d=0.69, [0.24, 1.13] | **d=0.78, [0.74, 0.83]** | **d=9.25, [8.67, 9.82]** |
| **Assimilation and contrast effects in question sequences** (Schwarz et al., 1991) | Subjects were asked "How satisfied are you with your relationship?" and "How satisfied are you with your life-as-a-whole?" in the two possible orders. | q=0.48, [0.07, 0.88] | q=−0.07, [−0.12, −0.02] | <u>q=0.06, [−0.07, 0.18]</u> |
| **How choosing vs. rejecting affects relative desirability** (Shafir, 1993) | Subjects chose whether to award custody or to deny custody to one of two parents: a parent of extreme characteristics (either strongly positive or strongly negative) and a parent of average characteristics. | d=0.35, [−0.04, 0.68] | d=−0.13, [−0.18, −0.09] | **d=2.11, [1.56, 2.67]** |
| **Perceived intentionality for side effects** (Knobe, 2003) | Subjects were asked whether a corporation vice president's decision to bring about a helpful or harmful side effect was intentional. The two were compared. | d=1.45, [0.79, 2.77] | **d=1.75, [1.70, 1.80]** | "correct answer" effect |

**Table 2**. Qualitative comparison between our GPT3.5 results, the Many Labs 2 results, and the original results, excluding the studies that were unanalysed for our GPT3.5 sample due to the "correct answer" effect of zero or near-zero variation in answers. The percentage of analysed studies whose results were successfully replicated—for each pair of samples—is listed below.

| | **Original effect** | **ML2 effect** | **GPT effect** | **ML2 replicates original** | **GPT replicates original** | **GPT replicates ML2** |
|---|---|---|---|---|---|---|
| **Structure promotes goal pursuit** (Kay et al., 2014) | Structured events are associated with higher willingness to pursue goals | Structured events are not associated with higher willingness to pursue goals | Structured events are not associated with higher willingness to pursue goals | No | No | Yes |
| **Consumerism undermines trust** (Bauer et al., 2012) | Consumer framing resulted in lower trust | Consumer framing resulted in lower trust | Consumer framing did not result in lower trust | Yes | No | No |
| **Disgust sensitivity predicts homophobia** (Inbar et al., 2009) | Actions are seen as more intentional for homosexual kissing than of heterosexual kissing | Actions are not seen as more intentional for homosexual kissing than of heterosexual kissing | Actions are seen as less intentional for homosexual kissing than of heterosexual kissing | No | No | No |
| **False consensus: traffic-ticket scenario** (Ross et al., 1977) | Choosing an option is associated with a higher estimation of frequency of this choice | Choosing an option is associated with a higher estimation of frequency of this choice | Choosing an option is associated with a higher estimation of frequency of this choice | Yes | Yes | Yes |
| **Reluctance to tempt fate** (Risen & Gilovich, 2008) | Likelihood estimations were higher when fate was tempted | Likelihood estimations were higher when fate was tempted | Likelihood estimations were lower when fate was tempted | Yes | No | No |
| **Less-is-better effect** (Hsee, 1998) | The higher-price cheap item is seen as more generous than the lower-price expensive item | The higher-price cheap item is seen as more generous than the lower-price expensive item | The higher-price cheap item is seen as more generous than the lower-price expensive item | Yes | Yes | Yes |
| **Assimilation and contrast effects in question sequences** (Schwarz et al., 1991) | Asking specific life satisfaction questions before general ones resulted in higher correlations | Asking specific life satisfaction questions before general ones resulted in lower correlations | Asking specific life satisfaction questions before general ones did not result in different correlations, though both relationships were negative | No | No | No |
| **Effect of choosing versus rejecting on relative desirability** (Shafir, 1993) | The extreme parent was both more likely to be awarded and be denied custody. | The extreme parent was both less likely to be awarded and be denied custody. | The extreme parent was both more likely to be awarded and be denied custody. | No | Yes | No |
| **Percentage replicated** | | | | **50%** | **37.5%** | **37.5%** |

## 3.1. The "correct answer" effect

Unexpectedly, we could not analyse six of the 14 studies in the manner we had originally planned in our pre-registration. In these six studies, different runs of GPT3.5 in our sample responded with zero or near-zero variation for either a dependent variable or condition variable question, in stark contrast to the significant variation shown by the corresponding human subjects. We call this the *"correct answer" effect*. This terminology denotes GPT3.5's tendency to sometimes answer survey questions—in a highly (or sometimes completely) uniform way. We take this pattern of responses to indicate that these LLM outputs are uniform because the LLM treats the question as if there was a correct answer. Of course, the questions we studied, touching on nuanced topics like political orientation, economic preference, judgement, and moral philosophy, do not lend themselves to correct answers, as can be seen in the diversity of opinions that human participants and subject-matter experts express about these issues. For the purposes of our analysis, we define a "correct answer" to be an answer given by 99% or more of surveyed GPT runs for a central-variable question, although this threshold is arbitrary.

One example of the "correct answer" effect was observed in the context of the Moral Foundations Theory survey of Graham et al. (2009), which probes political orientation and consequent moral reasoning. In this survey, subjects are asked to self-identify their political orientation. Then, self-identified liberals, moderates, and conservatives are asked to rate how relevant the concepts of harm, fairness, ingroup, authority, and purity (three survey questions per concept) are for deciding whether something is right or wrong, and their answers are compared. But in our GPT sample (N=1,030), we found that 99.6% of surveyed GPT3.5 runs (a total of 1,026) self-identified as a maximally strong conservative, while the remaining 0.4% of surveyed runs (a total of just four) all self-identified as political moderates. No GPT3.5 runs in our sample identified as any shade of political liberal, or indeed as any category of political orientation other than the two listed above. Because of the unexpected rarity of moderates and the complete lack of liberals in our planned GPT sample, our pre-registered analysis plan to compare the Moral Foundations of liberals and conservatives ended up being unsuitable.

Additionally, the survey of Rottenstreich and Hsee (2001) probes a certain economic preference. In it, subjects in one condition are asked to choose whether they would prefer a kiss from a favourite movie star versus $50. Subjects in the other condition are asked to make the same choice, but each outcome is awarded with 1% probability. We unexpectedly found in both conditions that 100% of surveyed GPT3.5 runs (N=1,040, with 520 in each condition) preferred the movie star's kiss. The uniformity of answers made the planned analysis infeasible. Due to the uniformity of answers and the consequent unconstructability of the statistical test we planned to run, we were unable to follow our pre-registered analysis plan. An illustrated comparison between the distribution of answers for the original sample of Rottenstreich and Hsee, the Many Labs 2 sample, and our GPT sample can be found in Figure 3.

Furthermore, the survey of Hauser et al. (2007) probes a question about moral philosophy. In it, subjects are asked whether various actions that sacrificed one person's life to save five people were morally permissible. The survey tests whether such a sacrifice would be less likely to be considered permissible if it is deemed as motivated by the greater good rather than as a foreseen side effect. The former was represented in a scenario where the focal individual pushed a large man in front of an incoming trolley to save five people's lives. The latter was represented in a scenario where the focal individual changed the trajectory of an out-of-control trolley so that it killed one person instead of five. Our sample of surveyed GPT3.5 runs (N=1,030) did show analysable variation in answers about the latter scenario's action, with 36% of the surveyed runs (total of 373) answering that it was morally permissible and 64% of them (total of 656) answering that it was not. On the other hand, the former scenario's action was regarded by 100% of the surveyed GPT3.5 runs as impermissible. This does directionally replicate the original finding of Hauser et al., but the unexpected uniformity of answers to the aforementioned survey question made the statistic we planned to analyse unconstructable, due to which we were technically unable to follow our pre-registered analysis plan.

**Figure 3**. Response distributions of whether subjects preferred a kiss from a favourite movie star or $50 when both outcomes were certain, left; and when both outcomes were awarded with 1% probability, right. The data pertains to the survey provided by the study of Rottenstreich and Hsee (2001) testing the relationship between affect and risk.

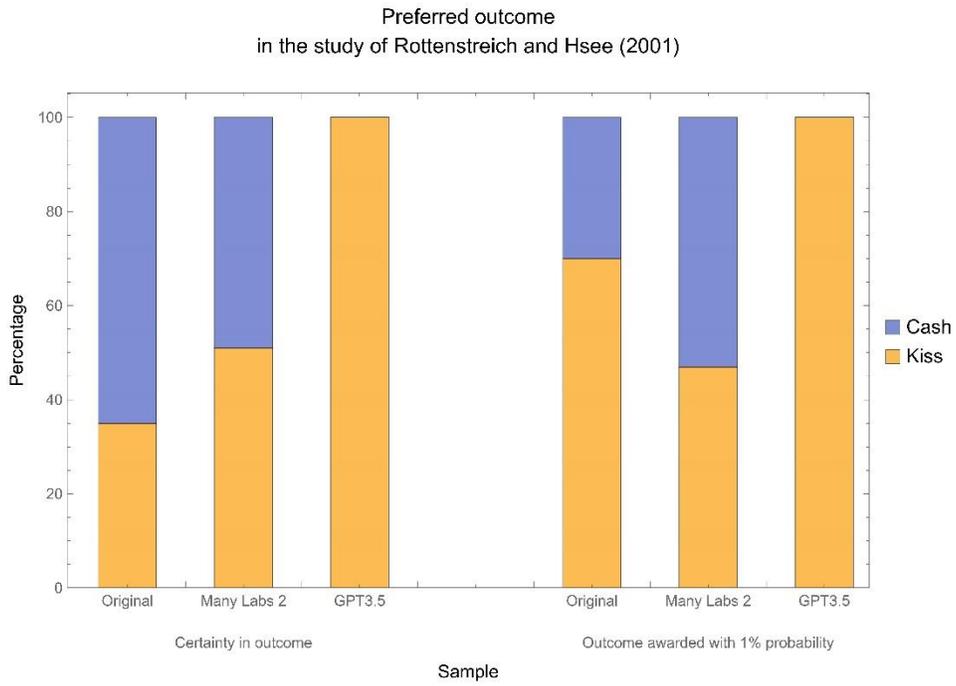

Additionally, the survey of Ross et al. (1977) probes both personal preference and judgement. Subjects are asked to estimate the probability that they would sign a release allowing footage that had recorded them to be used in a supermarket commercial, and to estimate others' probability of this action as well. The hypothesis was that subjects would be subject to the false-consensus belief: that their opinion will be more prevalent among others than it is in reality. But in our GPT sample (N=1,030), 99.7% of surveyed GPT3.5 runs (a total of 1,027) chose to sign the release, and only 0.3% of them (a total of just three) refused. This lack of variation in answers reduced the degrees of freedom for our pre-registered analysis plan, which thereby ended up being unsuitable.

The survey of Tversky and Kahneman (1981) again probes both personal preference and judgement. In it, subjects are divided into two conditions, each of which asks whether they would buy their desired items (one cheap and one expensive) from the store they are currently at or from a far-away branch of the store that sells one of the items for a lower price. In one condition, the cheap item is sold for a lower price; and in the other, the expensive item is sold for a lower price; but the cost saving for the two items combined is equal between the two conditions. Tversky and Kahneman's finding, replicated by the Many Labs 2 sample, was that people were more likely to travel to the far-away branch if the cost saving happens to be on the cheap item rather than the expensive item. However, we were unable to test this because of the complete uniformity of answers in our GPT sample (N=1,040, with 520 in each condition). All 100% of surveyed GPT3.5 runs in each condition chose to travel to the far-away branch of the store for the cost saving. Because the unexpected uniformity of answers made the statistic we planned to analyse unconstructable, we were once again unable to follow our pre-registered analysis plan.

And finally, the survey of Knobe (2003) investigates judgements of intentionality. In the study, subjects read a passage describing the decision of a company's board chairman that brought about either a harmful side effect or a helpful side effect—the two conditions—after which the subjects answer whether the board chairman intentionally brought about the side effect. The original finding was that the board chairman's action was more likely to be perceived as intentional if the side effect was negative. Many Labs 2 replicated this finding with a seven-point scale ranging from 'a lot of blame/praise' to 'no blame/praise', rather than a two-point scale ranging from intentional to unintentional. In our GPT sample (N=1,040, with 520 in each condition), the question with the seven-point scale showed "correct answers" for both conditions. Specifically, 99.2% of surveyed GPT runs (a total of 516) described the positive side effect as deserving of the highest degree of praise, or "a lot of praise"; 0.2% of them (a total of just one) described it as deserving of the second-highest degree of praise; and 0.6% of them (a total of 3) described it as deserving of the lowest degree of praise, or "no praise." Similarly, 100% of surveyed GPT runs described the negative side effect as deserving of the highest degree of blame, or "a lot of blame." Our pre-registered analysis plan was made unsuitable by the unexpected uniformity of GPT3.5's answers in the negative-side-effect condition.

While we have presently focused on GPT3.5's "correct answers" of near-complete or complete homogeneity, we note that the "correct answer" effect may also encompass GPT3.5's tendency to sometimes respond to different conditions of a given input with much more predeterminedly, but still no uniform, answers than did human subjects. To illustrate, consider GPT3.5's unprecedentedly large effect (d=9.25) in the same direction with the findings of Hsee (1998) and of Many Labs 2. Different runs of GPT3.5 thought the "correct answer" was that the higher-price variant of an inexpensive item (scarf) should be seen as more generous than lower-price variant of an expensive item (coat) to a much more predetermined degree than did human subjects.

**3.2. Exploratory robustness checks: Order effects and demographic prompt additions**

We conducted an exploratory follow-up study for each of the six studies where a central-variable question was answered with a homogeneous "correct answer." In each of our follow-up conditions, we presented the answer choices for the question with the homogeneous "correct answer" in the reverse order of our original condition to test for potential order effects. The results of these follow-up conditions are presented in Table 3. For the purposes of this analysis, we say that the "correct answer" was *robust to the order change* if 90% of surveyed GPT runs still gave the "correct answer" in the reverse-order condition. In summary, 66.7% of the analysed "correct answers," spanning six out of the nine studies, were robust to changing the order of answer choices. However, 33.3% of the "correct answers," spanning three out of the nine studies, were not robust to the order change.

For the study of Rottenstreich and Hsee (2001), we presented the option of a favourite movie star's kiss after rather than before the option of cash. For the condition in which outcomes were awarded with certainty, the same number of runs was surveyed (N=520), and the changed order of answer choices did not have an effect on GPT runs' responses. Just like for the original-order condition, 100% of GPT runs preferred the movie star's kiss in the reversed-order condition. For the condition in which outcomes were awarded with 1% probability, the same number of runs was surveyed (N=520), but the new order substantially changed the original finding that 100% of GPT runs preferred the movie star's kiss. In the reversed-order condition, 54% of the runs (total of 281) retained the original preference of the kiss, while 46% of them (total of 239 runs) preferred the cash. This order effect (Hohensinn & Baghaei, 2017) was much larger than ones that are typically seen in human subjects.

**Table 3.** Response distributions of the original sample, Many Labs 2 sample, our original-order GPT sample, and our reverse-order GPT sample for the ten considered "correct answers". Here, GPT runs' political orientations were categorised into "liberal," "moderate," and "conservative" so as to match the categories of Graham et al. (2009).

| | Original sample | ML2 sample | GPT sample (Original order) | GPT sample (Reverse order) | Robust to order change? |
|---|---|---|---|---|---|
| **Self-reported political orientation** (Graham et al., 2009) | 59% liberal<br>24% moderate<br>17% conservative<br>*(N=1,532)* | 38% liberal<br>39% moderate<br>23% conservative<br>*(N=6,966)* | 0% liberal<br><1% moderate<br>>99% conservative<br>*(N=1,030)* | >99% liberal<br><1% moderate<br>0% conservative<br>*(N=1,030)* | No |
| **Certain kiss versus certain cash** (Rottenstreich & Hsee, 2001) | 35% kiss<br>65% cash<br>*(N=20)* | 51% kiss<br>49% cash<br>*(N=3,493)* | 100% kiss<br>0% cash<br>*(N=520)* | 100% kiss<br>0% cash<br>*(N=520)* | Yes |
| **1% probability of kiss versus 1% probability of cash** (Rottenstreich & Hsee, 2001) | 70% kiss<br>30% cash<br>*(N=20)* | 47% kiss<br>53% cash<br>*(N=3,725)* | 100% kiss<br>0% cash<br>*(N=520)* | 54% kiss<br>46% cash<br>*(N=520)* | No |
| **Is pushing a large man in front of a trolley to save five people morally permissible?** (Hauser et al., 2007) | 11% permissible<br>89% impermissible<br>*(N=2,646)* | 17% permissible<br>83% impermissible<br>*(N=6,842)* | 0% permissible<br>100% impermissible<br>*(N=1,030)* | <1% permissible<br>>99% impermissible<br>*(N=1,030)* | Yes |
| **In the supermarket scenario, sign the release form for the video footage?** (Ross et al., 1977) | 66% sign<br>34% refuse<br>*(N=80)* | 54% sign<br>46% refuse<br>*(N=7,205)* | >99% sign<br><1% refuse<br>*(N=1,030)* | 92% sign<br>8% refuse<br>*(N=1,030)* | Yes |
| **The cheap item is discounted at a distant store. Go for the discount?** (Tversky & Kahneman, 1981) | 68% go<br>32% don't go<br>*(N=93)* | 49% go<br>51% don't go<br>*(N=3,609)* | 100% go<br>0% don't go<br>*(N=520)* | 100% go<br>0% don't go<br>*(N=520)* | Yes |
| **The expensive item is discounted at a distant store. Go for the discount?** (Tversky & Kahneman, 1981) | 29% go<br>71% don't go<br>*(N=88)* | 32% go<br>68% don't go<br>*(N=3,619)* | 100% go<br>0% don't go<br>*(N=520)* | >99% go<br><1% don't go<br>*(N=520)* | Yes |
| **Does the board chairman deserve blame for the harmful side effect?** (Knobe, 2003)<br><br>*degree one = no blame*<br>*degree seven = a lot of blame* | *n/a (due to using two-point scale instead of seven-point scale)* | <1% degree one<br>1% degree two<br>3% degree three<br>7% degree four<br>15% degree five<br>24% degree six<br>49% degree seven<br>*(N=4,000)* | 0% degree one<br>0% degree two<br>0% degree three<br>0% degree four<br>0% degree five<br>0% degree six<br>100% degree seven<br>*(N=520)* | 6% degree one<br>0% degree two<br>0% degree three<br>0% degree four<br>0% degree five<br><1% degree six<br>94% degree seven<br>*(N=520)* | Yes |
| **Does the board chairman deserve praise for the helpful side effect?** (Knobe, 2003)<br><br>*degree one = no praise*<br>*degree seven = a lot of praise* | *n/a (due to using two-point scale instead of seven-point scale)* | 18% degree one<br>12% degree two<br>7% degree three<br>7% degree four<br>4% degree five<br>2% degree six<br>1% degree seven<br>*(N=3,987)* | <1% degree one<br>0% degree two<br>0% degree three<br>0% degree four<br>0% degree five<br><1% degree six<br>>99% degree seven<br>*(N=520)* | 96% degree one<br><1% degree two<br>0% degree three<br>0% degree four<br>0% degree five<br>0% degree six<br>4% degree seven<br>*(N=520)* | No |
| **"Correct answers" replicated** | | | | | **66.7%** |

We also tested whether the two "correct answers" we observed for the study of Knobe (2003) were robust to order changes. One of the two "correct answers" did not replicate. In the question probing how much praise the board chairman deserves in the positive-side-effect condition, we presented the answer choices in reverse order, from 'A Lot of Praise' to 'No Praise' (N=520). This resulted in only 3.5% of surveyed GPT runs giving the original "correct answer" of 'A Lot of Praise' (a total of just 18 runs), 0.2% of GPT runs answering with the second-highest level of praise (a total of just one GPT run), and the remaining 96.3% of GPT runs answering with 'No Praise' (a total of 501 runs).

Finally, for the Moral Foundations Theory survey of Graham et al. (2009), we presented the options for self-reported political orientation in the order of "strongly conservative" to "strongly liberal," in contrast to the original order that ran in the other direction. The same number of runs were surveyed (N=1,030). The changed order of answer choices resulted in 99.3% of GPT runs self-identifying as "strongly liberal" (total of 1,023), in contrast to the 99.6% in the original condition self-identifying as "strongly conservative." In both conditions, GPT3.5 almost always self-identified as the political orientation given by the last presented choice.

Additionally, we conducted a second exploratory follow-up study—for the trolley-dilemma study of whether pushing a large man onto the tracks to save five others (Hauser et al. 2007)— in which we varied the demographic characteristics with which the LLM was prompted. This study aimed to test whether the lack of variation in some responses may be explained by a lack of demographic variation in the runs. For more information on the methods of this follow-up, see the Supplementary Information. In the study, we found that the "correct answer" effect persisted even when prompting the LLM to respond as a person with a random combination of demographic characteristics, such as a black 50-year-old Christian woman who has an advanced degree. Specifically, we replicated the "correct answer" effect of our original run, with 100% of GPT3.5 runs (N=982) indicating that it would be morally impermissible to shove a large man in front of a trolley. This suggests that at least some of the "correct answer" effects are not only insensitive to order effects, but are also insensitive to demographic variation in the prompt. This suggests that the "correct answer" effect may be relatively robust, and thus may surface in situations where LLM outputs may be used to supplant human decision-making.

### 3.3. Post-hoc rationalisation and right-leaning Moral Foundations

We conducted an unplanned exploratory follow-up analysis in which we computed our GPT runs' vector of average relevance values; and compared it with the vectors of the liberal subset (N=21,933), the moderate subset (N=3,203), the conservative subset (N=4,128), and the libertarian subset (N=2,999) among the human survey participants of Graham et al. (2011). We conducted the follow-up analysis for the self-reported GPT conservatives (N=1,026) that almost entirely comprised the original-order condition of our replication, and the self-reported GPT liberals (N=1,023) that almost entirely comprised the reverse-order condition. Table 4 presents the vectors of average relevance values reported by our samples of self-reported GPT liberals and self-reported GPT conservatives, the aforementioned human samples of Graham et al. (2011), and their comparisons. To compare these vectors, we used the absolute-error ($L^1$) distance metric, the Euclidean ($L^2$) distance metric, and the cosine similarity metric.

All three distance metrics found our sample of self-reported GPT conservatives to be most similar to conservative participants of Graham et al. (2011). This was always followed relatively closely by the moderate sample. The two furthest away were always the liberal sample (furthest away with respect to cosine similarity metric) and the libertarian sample (furthest away with respect to the $L^1$ and $L^2$ distance metrics).

When these three distance metrics were applied to our sample of self-reported GPT liberals, we found that first, according to each of the three distance metrics, self-reported GPT liberals had a lower distance to the human liberal sample and a higher distance to the human conservative sample than did the self-reported GPT conservatives. This was arguably an instance of post-hoc rationalisation, in which GPT3.5's answers to subsequent survey questions were chosen in a way that better fit its previous response. This post-hoc rationalisation effect is unsurprising, given that GPT3.5 has been trained to predict the sequence of words that is most likely to follow the preceding sequence of words.

The second, arguably more surprising finding was that according to each of the three distance metrics, our sample of self-reported GPT liberals were still closer to the human conservative sample than it was to the human liberal sample. Also, the $L^1$ distance metric found that self-reported GPT liberals were—among human liberals, human moderates, human conservatives, and human libertarians—closest in response to human conservatives. The $L^2$ distance metric and the cosine similarity metric instead found self-reported GPT liberals to be

**Table 4. (a)** The average vector of Moral Foundations Theory relevance values (mean unparenthesised, standard deviation parenthesised) for self-reported GPT liberals and self-reported GPT conservatives, as well as for human liberals, moderates, conservatives, and libertarians sampled by Graham et al. (2011). **(b)** How similar the average vector of self-reported GPT liberals is to those of the human samples with respect to the absolute-error ($L^1$) distance metric, the Euclidean ($L^2$) distance metric, and the cosine similarity metric. **(c)** How similar the average vector of self-reported GPT conservatives is to those of the human samples with respect to the three distance metrics.

(a)

| Concept | Liberal GPT sample (N=1,023) | Conservative GPT sample (N=1,026) | Liberals (N=21,933) | Moderates (N=3,203) | Conservatives (N=4,128) | Libertarians (N=2,999) |
|---|---|---|---|---|---|---|
| Harm | 4.02 (0.43) | 3.84 (0.48) | 3.93 (0.76) | 3.68 (0.84) | 3.48 (0.89) | 3.26 (1.03) |
| Fairness | 4.33 (0.26) | 4.26 (0.25) | 4.04 (0.67) | 3.77 (0.77) | 3.44 (0.87) | 3.66 (0.90) |
| Ingroup | 2.10 (0.88) | 2.24 (0.77) | 2.06 (0.94) | 2.56 (1.00) | 3.03 (1.02) | 2.16 (1.10) |
| Authority | 3.36 (0.61) | 3.34 (0.47) | 1.88 (0.86) | 2.37 (0.90) | 2.81 (0.91) | 1.71 (0.95) |
| Purity | 2.88 (1.08) | 3.07 (0.94) | 1.44 (0.94) | 2.09 (1.09) | 2.88 (1.11) | 1.31 (1.03) |

(b)

| Distance metric | Liberals (N=21,933) | Moderates (N=3,203) | Conservatives (N=4,128) | Libertarians (N=2,999) |
|---|---|---|---|---|
| | \multicolumn{4}{c}{Distance or similarity to liberal GPT sample *(bold denotes closest)*} | | | |
| $L^1$ distance | 3.339 | 3.147 | **2.920** | 4.717 |
| $L^2$ distance | 2.085 | **1.499** | 1.505 | 2.493 |
| Cosine similarity | 0.9713 | **0.9882** | 0.9830 | 0.9708 |

(c)

| Distance metric | Liberals (N=21,933) | Moderates (N=3,203) | Conservatives (N=4,128) | Libertarians (N=2,999) |
|---|---|---|---|---|
| | \multicolumn{4}{c}{Distance or similarity to conservative GPT sample *(bold denotes closest)*} | | | |
| $L^1$ distance | 3.581 | 2.934 | **2.704** | 4.660 |
| $L^2$ distance | 2.215 | 1.513 | **1.324** | 2.548 |
| Cosine similarity | 0.9649 | 0.9873 | **0.9874** | 0.9665 |

closest to human moderates—another manifestation of post-hoc rationalisation, via comparatively left-leaning responses—but human conservatives comprised a close second. Just like for self-reported GPT conservatives, the two human samples furthest away from the self-reported GPT liberals were always the liberal sample (furthest away from with respect to cosine similarity metric) and the libertarian sample (furthest away with respect to the $L^1$ and $L^2$ distance metrics). We thus robustly find that self-reported GPT liberals revealed right-leaning Moral Foundations: a right-leaning bias of lower magnitude, but a right-leaning bias nonetheless.

## 4. Discussion

### 4.1. Implications of the "correct answer" effect

Recent work by Grossman et al. (2023) has suggested applying LLMs to a wide variety of empirical social-science research, ranging from "supplant[ing] human participants for data collection" to drawing on them as "simulated participants" for hypothesis generation. Our data bolster the case that empirical findings on GPT3.5, as a rule of thumb, should not be assumed to generalise to the human case. There are at least three reasons. First, unlike the corresponding human subjects, different runs of GPT3.5 answered some nuanced questions—on nuanced topics like political orientation, economic preference, judgement, and moral philosophy—with as high or nearly as high a predeterminedness as humans would answer 2+2=4, which we termed the "correct answer" effect. Second, some of these "correct answers" showed drastic changes when answer choices were presented in the reverse order—to the point of having swings in response patterns that are clearly uncharacteristic of human responses. Third, GPT3.5 replicated just 37.5% of the original findings for the eight analysed studies, in contrast to the Many Labs 2 project's 50% replication rate for these studies. Such behavioural differences were arguably foreseeable, given that LLMs and humans constitute fundamentally different cognitive systems: with different architectures and potentially substantial differences in the various ways by which each of them has evolved, learned, or been trained to mechanistically process information (Shiffrin & Mitchell, 2023). Yet, given the anticipated rise in LLM capabilities (Hu et al., 2023) and their and other AI models' potential automation of much of human economic activity due to cost reasons (OpenAI, 2020), the psychologies of these models may be increasingly studied for their own sake: rather than as a purported window into studying the psychologies of humans.

More "correct answers" in LLM behavioural data have been documented since our study, such as GPT-4's tendency to describe software engineers almost exclusively as male (98% male pronouns, 1% female pronouns, and 1% other pronouns), even when 22% of software engineers are female; and to describe administrative assistants almost exclusively as female (98% female pronouns, 2% male pronouns), even when 11% of administrative assistants are male (Bubeck et al., 2023). We are unsure about the cause of such "correct answers" by OpenAI's models. One hypothesis is that the "correct answers" were learned from training data. Another hypothesis is that the "correct answers" may have resulted—either inadvertently or intentionally—from fine-tuning and reinforcement-learning selection pressures applied to the model. And a third hypothesis is that the "correct answers" may have occurred due to modifications imposed by OpenAI on the level of inputs and/or outputs, rather than of the model itself. Uncovering the true cause of a given "correct answer" may be possible in theory if one had access to closed-source information on the model in question. But in practice, given the black-box nature of LLMs, it is plausible that no

one—not even the creators of the model at OpenAI—understands the true cause of this phenomenon at this moment.

We found that one-third of our "correct answers" did not replicate when the answer choices to the survey questions were presented in reverse order. The precise replication failures suggest that when GPT3.5 makes decisions, the learned heuristics by which it does so may apply differently in different situations. To illustrate, we hypothesise that GPT3.5's "correct answers" for the 1% probability condition in the study of Rottenstreich and Hsee (2001) were partially due to a primacy effect favouring the first out of two listed answer choices: a heuristic that applied in the opposite direction when the answer choices were listed in reverse order. Also, we hypothesise that including a seven-point scale in a survey with questions using other scales tended to cause a recency bias, in which the last answer choice of a long list of seven answer choices is favoured. For example, GPT3.5 "correctly" answered the seven-point scale for self-reported political orientation with the last option rather than its actual political orientation, whatever that may mean. That GPT3.5 tended to show a recency effect rather than a primacy effect when given a long list of answer choices replicates a similar finding by Atkinson and Shiffrin (1968) on human subjects. This is consistent with Atkinson and Shiffrin's explanation—and with the overall theory of the availability heuristic (Tversky & Kahneman, 1973)—that while the last answer choice in a long list is disproportionately likely to be available in one's memory during their decision-making, the first item in a long list is more likely to be unavailable at the time of decision-making.

However, two-thirds of the considered "correct answers" were in fact robust to changes in the order of answer choices. Additionally, our follow-up study showed that even adding various randomly selected demographic details to the prompt did not change the "correct answer" effect in the study for which we tested this, suggesting that the effect is unlikely to have been caused by a lack of demographic information in our prompting process. We argue that at least some of these "correct answers" may be more likely to correspond to robustly learned biases (Mehrabi et al., 2021): biases that may conceptually influence the model's answers to a wide variety of realistic prompts, even if the model is asked to respond in a myriad of demographically varying roles. Robust biases may even be conceptually shared by different models with overlapping training data. A potential example of this is provided by the transphobic behaviour of Microsoft's chatbot Tay, which stated that "Caitlyn Jenner isn't a feminist, he is out to destroy the meaning of real women" (Alba, 2016); and of the GPT-3-based Seinfeld-simulation model "Nothing, Forever," which stated "I'm thinking about doing a bit about how being transgender is actually a mental illness" seven years after Tay (Rosenblatt, 2023). Other examples of robust AI biases include a prison-recidivism prediction model—used for screening decisions about pretrial release, sentencing, and parole—that predicted with higher false-positive rates for African-American individuals than Caucasian individuals (Angwin et al., 2016), a resume-screening model that learned to penalise women job applicants (Grossman, 2018), a beauty-contest-judging model that learned to penalise darker-skinned contestants (Levin, 2016), a facial-recognition model that overly mispredicted Asian individuals as blinking (Rose, 2010), and an advertisement model that underpromoted ads for Science, Technology, Engineering, and Mathematics (STEM) careers to young women (Lambrecht & Tucker, 2019).

The hypothetical AI models of the future may not only present information to numerous people as search-engine chatbots, AI assistants, and writers of media content; but may also plausibly automate other important roles in society (Ernst et al., 2019; Solaiman et al., 2019). These societally embedded AI models of the future may turn out to have learned from their training data certain predetermined characteristics of psychology, especially since much of the training data from which GPT3.5 may have learned its "correct answers" will also plausibly be used to train the hypothetical AI models of the future. There also remains the risk that current LLM output will itself be used as training data for further runs. This invites both a concern about the potential penalty to diversity of thought in such an AI-led future and a scientific desire to identify the highly predictable aspects and biases of AI psychologies. Future research that aims to predict whether AI systems will answer a given nuanced question with *(1)* a blunt, supposedly "correct answer" or *(2)* a non-predetermined answer more characteristic of human subjects would potentially be fruitful.

### 4.2. Implications of right-leaning Moral Foundations

We also unexpectedly found that the responses of both self-reported GPT liberals and self-reported GPT conservatives robustly lean towards the right. This result is in line with those of Abdulhai et al. (2023), who also used the Moral Foundation Theory survey to probe not just *text-davinci-003*, but also other models in the GPT-3 family (specifically, *text-davinci-002*, *text-curie-001*, and *text-babbage-001*), and found that the responses of every tested model were closest to those of conservative human subjects, suggesting that the results documented in our paper may generalise across models to a degree that could not be concluded from our results alone.

Why did GPT-3 models robustly reveal right-leaning Moral Foundations? We do not, and perhaps cannot, know for sure without access to closed-source information on the models. And even if OpenAI were to provide detailed information on the models' weights, training datasets, and training procedures, it may still be computationally difficult to identify the true cause of this phenomenon with high certainty. Our guess for the true cause is the largely internet-based training data, which we hypothesise to have a conservative bias when weighted with respect to factors (e.g., visibility, engagement) that increase their likelihood of being included in the training sets of LLMs, which then filters down through to the results obtained in this and similar studies.

The hypothesis that internet data has a *de facto* conservative bias encapsulates several other empirical phenomena, such as the tendency of Microsoft's chatbot Tay (Alba, 2016) and of the Seinfeld-simulation model "Nothing, Forever" (Rosenblatt, 2023) to adeptly learn anti-liberal behaviour and attitudes, the tendency of Google users' Search Engine Results Pages (SERPs) to be more right-leaning near the top than near the bottom (Robertson, 2018), the finding that the 40 most prominent websites in right-learning media have 2.7 times the total visibility on Google Search when compared to the 40 most prominent websites in left-leaning media (O'Toole, 2021), and the finding that conservative content outperforms liberal content in the context of Twitter's algorithmic recommendation system (Huszár et al., 2021). Future studies of whether the training data of GPT-3 was conservatively biased—and of whether this caused GPT-3 to reveal right-leaning Moral Foundations—would potentially be fruitful.

The conservative bias of GPT3.5's outputs that our study measured may also be specific to the context of the study's prompt, rather than a stable feature of the model. According to our data, GPT3.5 values all five of the moral foundations strongly. Liberals tend to highly value harm and fairness, while conservatives tend to highly value in-group, authority, and purity. In our study, GPT3.5 valued all five foundations at high levels. When asked about fairness and harm values that human liberals tend to care about strongly, GPT-3.5 answered that it strongly cares about these values; and when asked about purity, authority, and ingroup values that human conservatives tend to care about strongly, the model also answered that it strongly cares about these values. However, it is unclear whether the model would actually strongly exhibit these values in contexts other than that of the experiment, which solely prompted the model with survey questions asking whether they cared about each of these values. Future studies that probe the degree to which the conservative bias and strong value-adherence of GPT3.5's answers to the Moral Foundations questions (Graham et al., 2009) generalize to other prompt contexts would be fruitful in understanding the extent to which these results generalise.

### 4.3. Limitations

One limitation of our study is that its results may pertain only to GPT3.5 and not to other models, due to the causal psychological factors being potentially idiosyncratic to this model or the respective version of the model that we used. Different models will in general exhibit different behaviours that are sometimes also observed within models at different iterations. For example, a collection of survey questions—albeit ones that were different from ours—measured that a certain temporal version of ChatGPT was oriented towards pro-environmentalism and left-libertarianism (Hartmann et al., 2023). If future studies robustly find a certain psychological aspect of ChatGPT to be left-libertarian and the corresponding aspect of other less publicly used GPT-3 models to be right-leaning, our hypothesis for why this occurred would be one of the following. First, the fine-tuning and reinforcement-learning selection pressures that were applied to GPT3.5—before its public release as the chatbot ChatGPT—may have changed the political leanings of the model. Second, the modifications added at the level of inputs and/or outputs may have caused the change. Each of these hypotheses is consistent with the finding that the strength and direction of ChatGPT's political leanings—as measured by certain survey questions—varied over time (Rozado, 2023), as OpenAI used public feedback on the chatbot to make closed-source updates between the release dates of the chatbot's multiple versions.

In addition, the capabilities of LLMs and of AI models in general will plausibly continue to grow at a fast rate. Thus, it is also possible that the hypothetically more powerful and emergently different models of the future may learn different psychological characteristics than did GPT3.5: even if much of GPT3.5's training data could plausibly also be used to train these hypothetical future models.

Another limitation of our study is that our method, the psychology survey, is an entirely text-based attitudinal response method. This limits the external validity of the findings compared to more behavioural choices that agents may make with their resources and time given agency over those. For instance, understanding the model at a deeper level than that of responses to surveys may be required for resolving key dilemmas, such as whether GPT3.5's "correct answers," its right-leaning Moral Foundations, its successful post-hoc rationalisation, and its human-like responses to various prompts reflect genuinely learned concepts rather than surface-level memorization from the relevant training data, and whether these are likely to impact actions when these systems are applied. This knowledge may be necessary for precisely predicting behaviour in unprecedented situations for which there is currently no data: although it should be noted that in practice, precision often comes at the expense of generality (Matthewson & Weisberg, 2009). Methods that have shown promise for studying human cognition at the mechanistic level rather than at the behavioural level—such as neuroscience and computational modelling—may also be promising for analogously studying a wide variety of AI cognitive systems, although such mechanistic studies may require access to closed-source information on the systems in question.

A further limitation of our study is that we have not replicated the studies of Many Labs 2 that drew on graphical information, which led to our subsample of the studies being non-random. For the present paper, we have chosen to replicate only a pre-registered implementation of straightforwardly automatisable surveys and questionnaires and did not set out to use trial-and-error to represent graphical information either as a text-based graphic or a prompt that communicated the necessary information. We believe, however, that communicating graphical information to GPT3.5 is very doable. Future research on the nascent field of prompt engineering—on how to effectively communicate to LLMs various forms of information, including but not limited to graphical information—would potentially be fruitful in expanding our results to these contexts.

## Code availability

Python 3.9.15 was used to collect survey data from OpenAI's *text-davinci-003* model, to convert the resulting .txt output files to .csv, and to create Figure S5 illustrating the Cohen's d effect sizes of several of our studies. The analyses were done primarily via the R-based GUI 'JAMOVI' and Excel; exceptions are detailed in the Supplementary Information. Figures 1 and 2 were made via Mathematica 13. The corresponding code for the above and other research steps is available online at the pre-registered OSF database (Park et al., 2023).

## Data availability

Our survey data, the primarily JAMOVI-based analyses of the data, the spreadsheet of Cohen's d effect sizes and 95% confidence intervals (used for Figure S5), and other relevant data are available at the pre-registered OSF database (Park et al., 2023).

## Acknowledgements

P.S.P. is grateful to be funded by the MIT Department of Physics and the Beneficial AI Foundation. The research views presented in this paper are solely the authors' and do not express the views of MIT, the institution of P.S.P.; of the London School of Economics, the institution of P.S.; or of CVS Health, the employer of C.Z., We are grateful to Maximilian Maier for his valuable contribution to the pre-registration writeup. We are also grateful to Mohammad Atari, Stephen Fowler, Ben Grodeck, Joe Henrich, Kevin Hong, Liav Koren, Ivan Kroupin, Raimund Pils, Konstantin Pilz, Slava Savitskiy, and Marc Wong for their helpful comments on the draft.

## Author contributions

P.S.P. and P.S. designed the research. P.S.P., P.S., and C.Z. performed the research. P.S.P. and P.S. wrote the paper.

## Open practices statement

The materials, data, and analysis files for all experiments are available at the pre-registered OSF database (Park et al., 2023). Most experiments were pre-registered, with the exception of the exploratory analyses and experiments inspired by the unexpected "correct answer" effect, which are indicated as such in the manuscript.

Henrich, J., Heine, S. J. & Norenzayan, A. (2010). The weirdest people in the world? *Behavioral and Brain Sciences*, 33(2-3), 61–83. https://doi.org/10.1017/S0140525X0999152X

Hohensinn, C. & Baghaei, P. (2017). Does the position of response options in multiple-choice tests matter? *Psicológica*, 38(1), 93.

Horton, J. J. (2023). Large Language Models as simulated economic agents: What can we learn from homo silicus? *arXiv* preprint arXiv:2301.07543.

Hsee, C. K. (1998). Less is better: When low-value options are valued more highly than high-value options. *Journal of Behavioral Decision Making*, 11(2), 107-121. https://doi.org/10.1002/(SICI)1099-0771(199806)11:2%3C107::AID-BDM292%3E3.0.CO;2-Y

Hu, L., Habernal, I., Shen, L., & Wang, D. (2023). Differentially private natural language models: Recent advances and future directions. *arXiv* preprint arXiv:2301.09112.

Huszár, F., Ktena, S. I., O'Brien, C., Belli, L., Schlaikjer, A., & Hardt, M. (2022). Algorithmic amplification of politics on Twitter. *Proceedings of the National Academy of Sciences*, 119(1), 1. https://doi.org/10.1073/pnas.2025334119

Inbar, Y., Pizarro, D. A., Knobe, J., & Bloom, P. (2009). Disgust sensitivity predicts intuitive disapproval of gays. *Emotion*, 9(3), 435-439. https://doi.org/10.1037/a0015960

John, O. P., & Srivastava, S. (1999). The Big-Five trait taxonomy: History, measurement, and theoretical perspectives. In L. A. Pervin & O. P. John (Eds.), *Handbook of Personality: Theory and Research* (pp. 102–138). Guilford Press.

Lambrecht, & Tucker, C. (2019). Algorithmic bias? An empirical study of apparent gender-based discrimination in the display of STEM career ads. *Management Science*, 65(7), 2966–2981. https://doi.org/10.1287/mnsc.2018.3093

Levin, S. (2016, September 8). *A beauty contest was judged by AI and the robots didn't like dark skin*. The Guardian. Retrieved March 8, 2023, from https://www.theguardian.com/technology/2016/sep/08/artificial-intelligence-beauty-contest-doesnt-like-black-people

Li, X., Li, Y., Liu, L., Bing, L., & Joty, S. (2022). Is GPT-3 a psychopath? Evaluating Large Language Models from a psychological perspective. *arXiv* preprint arXiv:2212.10529.

Majid, A (2023). Establishing psychological universals. *Nature Reviews Psychology*. https://doi.org/10.1038/s44159-023-00169-w

Matthewson, J. & Weisberg, M. (2009). The structure of tradeoffs in model building. *Synthese*, 170(1), 169–190. https://doi.org/10.1007/s11229-008-9366-y

Mehrabi, N., Morstatter, F., Saxena, N., Lerman, K., & Galstyan, A. (2021). A survey on bias and fairness in machine learning. *ACM Computing Surveys*, 54(6), 1–35. https://doi.org/10.1145/3457607

Metz, C. (2020, November 24). *Meet GPT-3. It has learned to code (and blog and argue)*. New York Times. Retrieved February 3, 2023, from https://www.nytimes.com/2020/11/24/science/artificial-intelligence-ai-gpt3.html

Miotto, M., Rossberg, N., & Kleinberg, B. (2022). Who is GPT-3? An exploration of personality, values and demographics. In *Proceedings of the Fifth Workshop on Natural Language*

# Supplementary Information

# Diminished Diversity-of-Thought in a Standard Large Language Model

Peter S. Park[1, *], Philipp Schoenegger[2, *], Chongyang Zhu[3]

[1] Department of Physics, MIT, 70 Vassar St., Cambridge, MA, USA (dr_park@mit.edu)

[2] Department of Management, London School of Economics, Marshall Building, 44 Lincoln's Inn Fields, London, England, UK (contact.schoenegger@gmail.com)

[3] CVS Health, 1 CVS Dr., Woonsocket, RI, USA (cyzhu95@gmail.com)

[*] Co-first author

## S.1. Methods

We pre-registered this study on the Open Science Framework (Park et al., 2023). For our study, we drew on the set of studies used in Many Labs 2 (Klein et al., 2018). For the entirety of this paper, we will continue to use Many Labs 2's numbering system of studies (Klein et al., 2018, 453–467) to allow for easy comparisons and reduce confusion risks. The total number of potential studies was 28. We excluded a total of 14 studies prior to data collection as they included pictures, compared national samples, relied on handwriting or font changes, or used an otherwise inapplicable component that were not transferable to a GPT context at this moment or without further prompt engineering. As such, we collected data for a total of 14 studies.

To collect our data, we called for OpenAI's *text-davinci-003* model (colloquially known as GPT3.5) to answer survey questions as a human subject of a social psychology experiment would. The parameters, detailed in OpenAI's API reference (OpenAI, 2023a), were set to the following: The number of maximum tokens per run was set to 2048 (max_tokens=2048). Note that tokens are "pieces of words, where 1,000 tokens is about 750 words" (OpenAI, 2023b). Stop sequences were not used (stop=None). The "temperature" parameter, which parametrizes whether outputs are more random or more deterministic, was set to the default value 1.0 (temperature=1.0). We set each call to GPT3.5 to result in 10 runs (n=10). The survey was put in string form (with "\n" denoting line breaks) as required by GPT3.5. As a prompt-engineering technique, we put before the survey questions instructions to GPT3.5 on how to format its output, and after the survey questions a "CHECKLIST FOR SURVEY ANSWERS" section to remind GPT3.5 of these formatting instructions. Due to these and other differences, the GPT experiments were not the same as those of the original survey or of the Many Labs 2 replication.

In order to obtain approximately 1,000 runs per study while accounting for the possibility of incorrect fill-outs of the survey, we collected slightly more runs. Specifically, we collected 520 runs per condition for the two-condition studies for a total of 1,040 runs (Studies 2, 5, 6, 8, 16, 18, 21, 24, 25, and 27). For the one-condition studies, we obtained either 1,020 runs (Study 14) or 1,030 runs (Studies 4, 11 and 13). Some studies' runs required multiple calls to OpenAI due to factors like accidental computer shutdown, internet shutdown, or typos in the inputted survey. For example, we found out that the first question of Study 4 (self-identification of political orientation) was erroneously left out, and that we had initially analysed the erroneous data from the fifteen-question version of the survey rather than the planned sixteen-question version. Also, we found out that the survey text strings for Study 24 had erroneously indicated in the "CHECKLIST FOR SURVEY ANSWERS" section that the total number of questions was three instead of the correct number two, leading to an unintended and factually inconsistent survey text. This led us to re-run getting the data for Studies 4 and 24 with the corrected survey text. The python code for surveying

GPT3.5 can be found on our Open Science Foundation (OSF) database at the link https://osf.io/dzp8t/?view_only=45fff3953884443d81b628cdd5d50f7a (Park et al., 2023).

We then converted the survey responses of GPT3.5 runs, which are originally in .txt format, to .csv for statistical analysis. All statistical analysis was conducted as done in the Many Labs 2 paper (Klein et al., 2018). We threw out survey responses of runs that did not answer all survey questions, or that answered survey questions incorrectly, such as responses that included characters that were not among the possible response categories. Details of our criterion for which runs' survey responses are thrown out can be found in the Python code on the OSF. Six of these 14 studies produced responses that had near-zero or zero variation in answers at all, meaning we are unable to analyse them statistically in the pre-registered way. This left us with a total of eight studies to analyse in line with our pre-registration plan, although we report some descriptives for the remaining six studies. Analyses were mostly conducted in the R-based (R Core Team, 2021) GUI 'JAMOVI' (The jamovi project, 2022) and in Excel. The relevant analysis files are also available at the pre-registered OSF database (Park et al., 2023).

## S.2. Results

For our 14 studies, we collected a total of 14,276 runs after excluding runs with missing data. There was substantial heterogeneity with respect to replication of effects, with some of our studies failing to show an effect, others replicating the effect in the same direction, and yet others demonstrating an effect in the opposite direction. In this section, we will outline the results for the eight studies individually. Then, we will quickly discuss the results for the six studies we could not analyse as planned in our pre-registration.

*STUDY 2.* Kay et al. (2014) surveyed 67 individuals to study the impact that structure in unrelated domains had on one's willingness to pursue goals. Their original sample found that those in the structural event condition (M=5.26, SD=0.88) were more likely to indicate willingness to pursue their goal compared to those in the random event condition (M=4.72, SD=1.32), t(65)=2.00, p=0.05, d=0.49, 95% CI=[0.001, 0.973]. The Many Labs 2 sample (N=6,506) failed to replicate this effect; those in the structural event condition (M=5.48, SD=1.45) were not significantly more likely to indicate willingness to pursue their goal compared to those in the random event condition (M=5.51, SD=1.39), t(6498.63)=-0.94, p=0.35, d=-0.02, 95% CI=[-0.07, 0.03]. Across our sample (N=992), GPT3.5 runs which were exposed to the structured event (M=6.18, SD=1.25) were not significantly more or less interested in pursuing their goals than those that were exposed to a random event (M=6.09, SD=1.25), t(990)=1.05, p=0.293, d=0.06, 95% CI=[-0.06, 0.19]. Thus, our GPT sample showed the same pattern of results as Many Labs 2, in that we also fail to find an effect.

*STUDY 4.* Graham et al. (2009) surveyed 1,548 individuals spanning the political spectrum to test how liberals and conservatives differ in how much they think the five concepts of Moral Foundations Theory—harm, fairness, ingroup, authority, and purity—are relevant for moral decision-making. They did so by asking subjects to self-identify on the political spectrum (on a seven-point scale ranging from "strongly liberal" to "strongly conservative," with the midpoint labelled "moderate"), and then asking about the relevance of each concept in the form of three different survey questions, for a total of 15 questions which were presented in random order. Graham et al. found that the individualising foundations of harm and fairness were more likely to be rated as relevant by liberals (r=−0.21, d=−0.43, 95% CI=[−0.55, −0.32]), while the binding foundations of ingroup, authority, and purity were more likely to be rated as relevant by conservatives (r=0.25, d=0.52, 95% CI=[0.40, 0.63]). The Many Labs 2 sample (N=6,966) tested whether the binding foundations were more likely to be rated as relevant by conservatives, a finding that they successfully replicated (r=0.14, p=6.05e−34, d=0.29, 95% CI=[0.25, 0.34], q=0.15, 95% CI=[0.12, 0.17]). We replicated this study, with the caveat that we did not randomise the order of questions, because to collect highly controlled and statistically valid data for all 15! ≈ 1.31e$^{12}$ possible conditions randomising the 15 survey questions—or even for all 5!=120 conditions randomising the five concepts—would have been impractical. We encountered the black-swan event in which across our GPT sample (N=1,030), an overwhelming majority of 99.6% of runs (total of 1,026) self-reported as a strong conservative, and the remaining 0.4% of runs (total of just four) self-reported as a moderate, with no shades of liberal in our sample. One additional "correct answer" we found pertained not to a central variable of analysis, but a component of a central variable. A surveyed GPT run's answer to "Whether or not someone was denied his or her rights" is one of three relevance values that goes into the relevance value of fairness. An overwhelming >99.9% surveyed runs (total of 1,029) answered that this was "always relevant": the maximal choice of answer on the six-point scale. The remaining <0.1% of runs (total of just one) answered that this was "very relevant": the second highest choice of answer. While our analysis plan would have been doable if this was the only "correct answer" in the data—as the focal item is only one of three items that averages out to the analysable fairness relevance value—the "correct answer" of maximal political conservatism made our analysis plan for comparing the responses of liberals and conservatives unsuitable.

Moreover, we conducted an unplanned study with the order of presented answers switched, so that "strongly liberal" was last rather than first. In the reverse-order sample, an overwhelming majority of 99.3% of runs (total of 1,023) self-reported as a strong liberal, and the remaining 0.7% of runs (total of just seven) self-reported as a moderate, with no shades of conservative. See Figure S1 for a visual summary.

*STUDY 5.* Rottenstreich and Hsee (2001) surveyed 40 individuals whether they would prefer a kiss from a favourite movie star (the *affectively attractive* option) or $50 (the financial option). Subjects in one condition made this choice, while subjects in the other condition made the same choice with the caveat that each option is awarded with 1% probability. In the probabilistic-outcome condition, 70% of the individuals preferred the movie star's kiss; whereas in the certain-outcome condition, only 35% preferred the kiss. The difference between the two conditions was significant, $\chi 2(1, N=40)=4.91$, $p=0.0267$, $d=0.74$, 95% CI=[< 0.001, 1.74]. In the Many Labs 2 sample (N=7,218), the probabilistic-outcome condition found 47% of individuals to prefer the movie star's kiss; whereas the certain-outcome condition found 51% to prefer the kiss. The effect was much smaller than and in the opposite direction of the original finding, but it was significant ($p=0.002$, OR=0.87, $d=-0.08$, 95% CI=[−0.13, −0.03]). We encountered the black-swan event in which across our GPT sample (N=1,040, with 520 in each condition), all 520 runs in each condition preferred the kiss, regardless of whether it was certain or probabilistic. This made our pre-registered analysis plan impossible, in that the statistic we had planned to analyse could not be even constructed in a well-defined manner.

We conducted an unplanned study with the order of presented answers switched (N=1,040, with 520 in each condition). However, this only created variation away from the kiss preference answer in the probabilistic-outcome condition, with 54% of runs (total of 281) preferring the 1% probability of a kiss and 46% of them (total of 239 runs) preferring the 1% probability of the money. The certain-outcome condition still uniformly preferred the kiss. See Figure S2 for a visual summary.

**Figure S1**. Subjects' self-reported political orientation for the Moral Foundations Theory survey of Graham et al. (1977). Fine-grained responses on the seven-point scale were binned into "liberal," "moderate," and "conservative" for inter-sample comparability. The "correct answer" was given by the last presented answer choice: "strongly conservative" in the original-order condition and "strongly liberal" in the reverse-order condition.

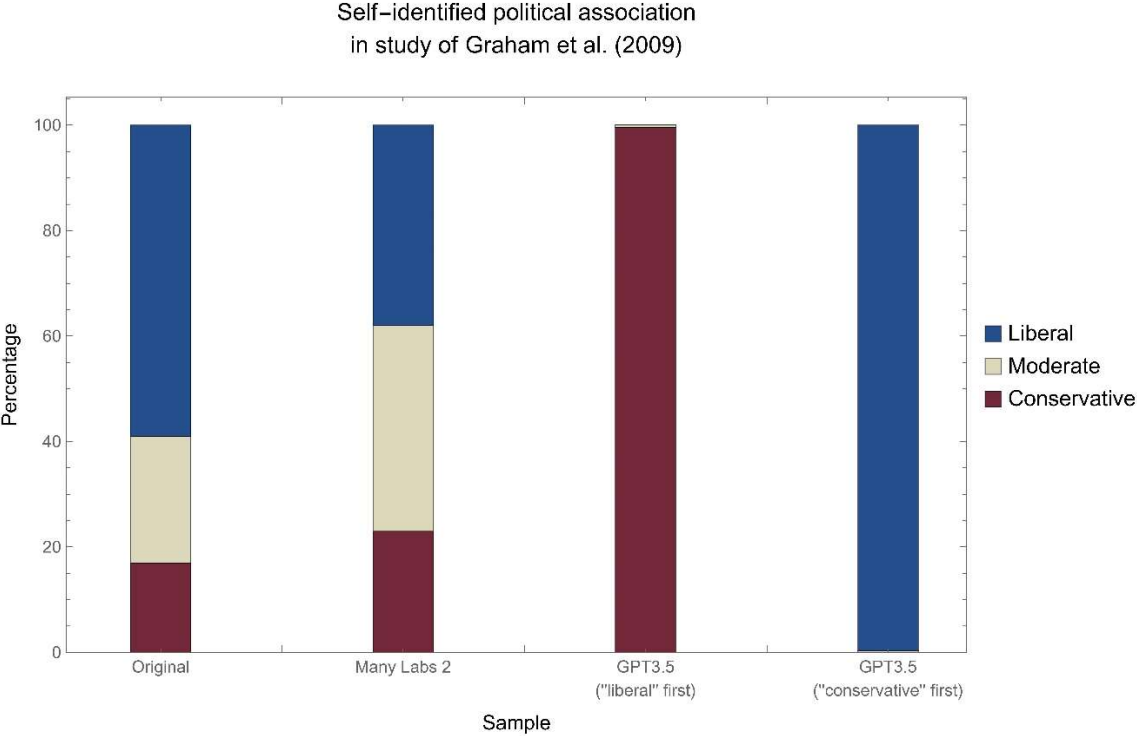

**Figure S2**. Subjects' answers when asked, for the survey of Hauser et al. (2007), whether pushing a large man in front of an incoming trolley to save five people was morally permissible. The "correct answer" of responding that the action is morally impermissible was robust to reversing the order of answer choices.

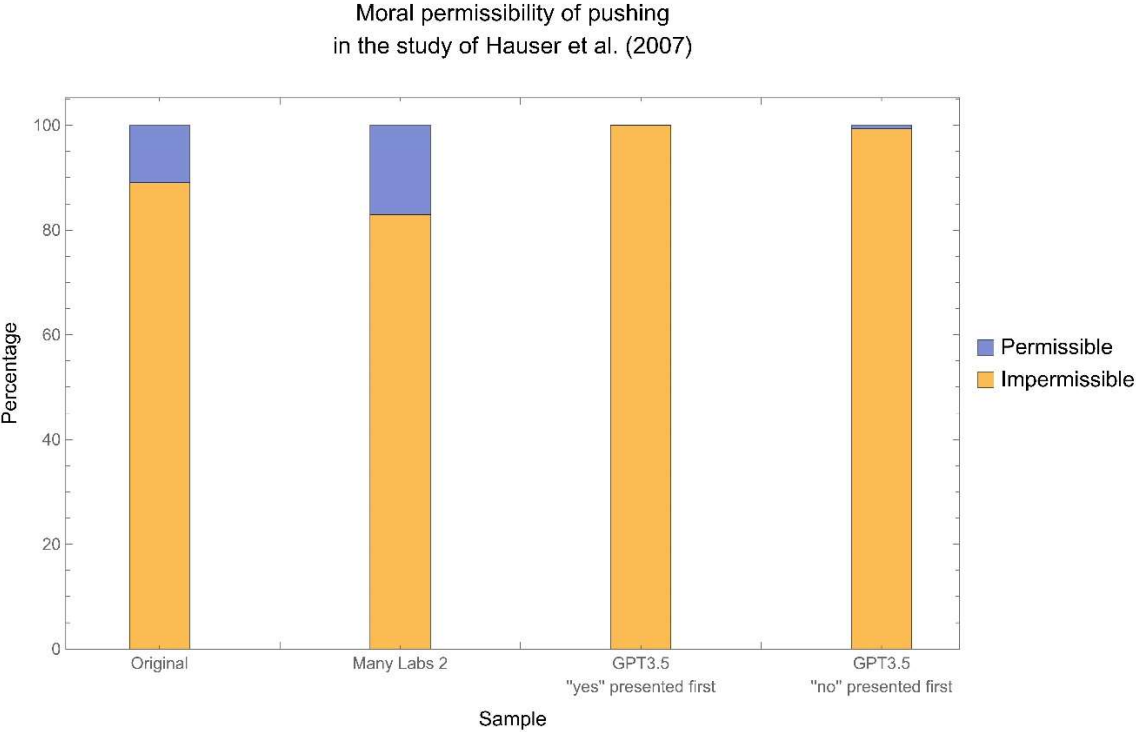

*STUDY 6.* Bauer et al. (2012) surveyed 77 individuals to look at the relationship between consumer mindsets and trust in others in a water conservation scenario. Their original sample found that referring to others as "consumers" (M=4.08, SD=1.56; here, 1 denotes "not at all" and 7, "very much") resulted in lower trust—that others would conserve water—than referring to them as "individuals" (M=5.33, SD=1.30), t(76)=3.86, p=0.001, d=0.87, 95% CI=[0.41, 1.34]). The Many Labs 2 sample (N=6,608) replicated this original effect, although its effect was markedly smaller; those in the "consumer" condition (M=3.92, SD=1.44) reported lower trust in a water conservation scenario than those in the control "individual" condition (M=4.10, SD=1.45), t(6606)=4.93, p=8.62e$^{-7}$, d=0.12, 95% CI=[0.07, 0.17]. In our sample (N=1,040), GPT3.5 runs in the "consumer" condition (M=3.40, SD=0.60) did not show a significantly different response than GPT3.5 runs in the control "individual" condition (M=3.34, SD=0.61), t(1038)=1.73, p=0.083, d=0.11, 95% CI=[-0.01, 0.23]. As such, we fail to replicate this effect.

*STUDY 8*. In the study of 44 individuals by Inbar et al. (2009), participants judged a director's action of portraying homosexual kissing as more intentional (M=4.36, SD=1.51) than that of portraying heterosexual kissing (M=2.91, SD=2.01), β=0.41, t(39)=3.39, p=0.002, r=0.48. The correlation between disgust sensitivity and judgement of intentionality was positive in the homosexual kissing condition, β=0.79, t(19)=4.49, p=0.0003, r=0.72; and negative in the heterosexual kissing condition, β=-0.20, t(19)=-0.88, p=0.38, r=0.20. The former correlation was stronger than the latter, z=2.11, p=0.03, q=0.70, 95% CI=[0.05, 1.36]. The Many Labs 2 sample (N=7,117) failed to find this effect on intentionality. Participants did not judge a director's action of portraying homosexual kissing as more intentional (M=3.48, SD=1.87) than that of portraying heterosexual kissing (M=3.51, SD=1.84), t(7115)=-0.74, p=0.457, d=-0.02, 95% CI=[-0.06, 0.03]. Disgust sensitivity and judgement of intentionality were positively related in both the homosexual kissing condition, r=0.12, p<0.001; and the heterosexual kissing condition, r=0.07, p<0.001. The correlation in the homosexual kissing condition and that in the heterosexual kissing condition were similar, z=2.62, p=0.02, q=0.05, 95% CI=[0.01, 0.10]. In our sample (N=1,040), GPT3.5 runs judged the director's action as less intentional in the context of homosexual kissing (M=4.45, SD=2.14) than in the heterosexual kissing condition (M=5.59, SD=1.79), t(1008)=-9.35, p<0.001, d=-0.580, 95% CI=[-0.71, -0.45]. We also find that the relationship between judgements of intentionality and disgust sensitivity was present but negative in the homosexual kissing condition, r=-0.13, p=0.003, whereas we did not find such a relationship in the control, r=-0.04, p=.365. Our GPT study found an effect in the opposite direction compared to the original effect, whereas Many Labs 2 had not found any effect.

*STUDY 11*. In the study of Hauser et al. (2007), N=2,646 individuals were asked about two trolley dilemmas. In the *foreseen-side-effect* scenario, the focal individual changes the trajectory of an out-of-control trolley to kill one person instead of five people. In the *greater-good* scenario, the focal individual pushes a large man off a bridge in front of an incoming trolley to stop it and thereby save five people's lives. They found that 89% of subjects deemed the action in the foreseen-side-effect scenario as permissible (95% CI=[87%, 91%]), while only 11% of them deemed the action in the greater-good scenario as permissible (95% CI=[9%, 13%]). The difference between the two percentage values was significant, $\chi^2$(1, N=2,646)=1,615.96, p<0.001, w=.78, d=2.50, 95% CI=[2.22, 2.86]. The Many Labs 2 sample (N=6,842) successfully replicated this finding. 71% of subjects deemed the action in the foreseen-side-effect scenario as permissible, while only 17% of them deemed the action in the greater-good scenario as permissible. The difference between the two percentage values was significant, p=2.2$e^{-16}$, OR=11.54, d=1.35, 95% CI=[1.28, 1.41]. In our GPT sample (N=1,030) the foreseen-side-effect scenario's action was deemed permissible by 36% of surveyed runs (total of 373) and impermissible by 64% of them (total of 656). However, the greater-good scenario's action was deemed impermissible by all 100% of surveyed runs. While this does successfully replicate the original finding of Hauser et al. and the Many Labs 2 finding, the unexpected uniformity of answers in the greater-good scenario's central variable made the statistic we planned to analyse unable to be constructed in a well-defined manner, due to which we were technically unable to follow our pre-registered analysis plan.

We conducted an unplanned study with the order of presented answers switched for the question pertaining to the aforementioned scenario (N=1,030). However, the "correct answer" was robust to this order change. An overwhelming 99.3% of surveyed GPT Runs (a total of 1023) still responded that pushing the large man to save five people was impermissible, whereas only 0.7% of GPT runs (a total of just seven) responded that it was permissible.

*STUDY 13*. Ross et al. (1977) provided early evidence for the false consensus effect, which shows that people's estimates of the frequency of any given belief is biased towards that person's own beliefs. In their study, 320 participants were presented with one of four hypothetical events (one of which was the supermarket scenario) and a corresponding choice between two action options. Those who chose the first option—compared to those who chose the second—estimated that a higher percentage of the other participants would choose the first option (M=65.7% vs. 48.5%), $F(1, 312)=49.1$, $p<0.001$, $d=0.79$, 95% CI=[0.56, 1.02]. The Many Labs 2 sample (N=7,205) replicated the supermarket scenario of the original study. Its results provided evidence in favour of replication, finding that those choosing the first option also believed that a higher percentage of people would choose that option (M=69.19% vs. 43.35%), $t(6420.77)=49.93$, $p<0.001$, $d=1.18$, 95% CI=[1.13, 1.23]. In our sample (N=1,030), however, 99.7% of surveyed runs (a total of 1,027) answered they would sign the release, while only 0.3% of them (a total of just three) answered they would not. The uniformity in answers reduced the degrees of freedom for our pre-registered analysis plan, making it unsuitable.

We conducted an unplanned study with the order of presented answers reversed for the question pertaining to the aforementioned scenario (N=1,030). However, the "correct answer" was robust to this order change. An overwhelming 92.0% of surveyed GPT Runs (a total of 948) still responded that they would agree to sign the release agreement, whereas only 8.0% of GPT runs (a total of 82) responded that they would refuse. See Figure S3 for a visual summary.

*STUDY 14*. Ross et al. (1977) also examined the false consensus effect in the context of a traffic-ticket scenario, finding evidence for the same effect as above, $F(1, 78)=12.8$, $d=0.80$, 95% CI=[0.22, 1.87]. As before, the Many Labs 2 sample for the traffic-ticket scenario showed that the effect was replicated; those choosing the first option (N=7,827) also believed that a higher percentage of people would choose that option, (M=72.48% vs. 48.76%), $t(6728.25)=41.74$, $p<0.001$, $d=0.95$, 95% CI=[0.90, 1.00]. In our GPT sample (N=1,020), we also find that those who reported willingness to pay the fine believed that a higher percentage of others would pay the fine compared to those who chose to go to court, (M=73.1% vs. M=58.7%, $t(1018)=18.0$, $p<0.001$, $d=1.27$, 95% CI=[1.11, 1.42]. This pattern of data suggests that our results also replicated the original effect.

*STUDY 16*. In the study of Tversky & Kahneman (1981) on the effect of framing on decision-making, 181 individuals were surveyed about a situation where they were tasked with buying two items: a cheap item (priced at $15) and an expensive item (priced at $125) at a store. 93 of the individuals were assigned to the condition where the cheap item could be purchased for $5 less at the store's other branch, a 20-minute drive away. In the sample, 88 of the individuals

**Figure S3**. Subjects' answers when asked, for the supermarket scenario in the study of Ross et al. (1977), whether they would sign a release agreement for video footage on them to be used for a supermarket commercial. The "correct answer" of signing the release was robust to reversing the order of answer choices.

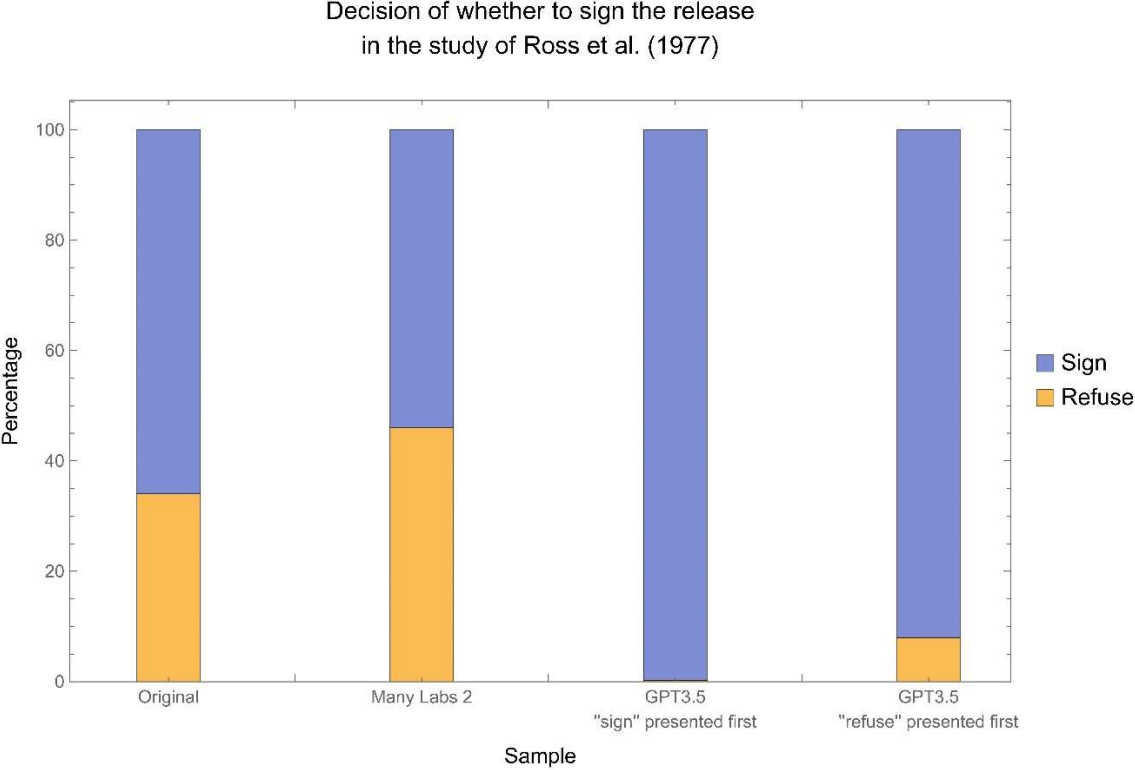

were assigned to the condition where the expensive item could be purchased for the same amount less at the other store. Consequently, the hypothetical cost saving was equal in both conditions. Individuals tended to decide to go to the other store more when the cost saving applied to the cheap item (68%) than when it applied to the expensive item (29%). This difference was statistically significant ($z=5.14$, $p=7.4e^{-7}$, OR=4.96, 95% CI=[2.55, 9.90]), resulting in the finding that the decision of whether to go to the distant store was affected by the base cost of the discounted item rather than just the total discount. The Many Labs 2 sample (N=7,228) replicated this with inflation-adjusted prices and different item types. They successfully replicated Tversky and Kahneman's finding, in that individuals tended to decide to go to the other store more when the cost saving applied to the cheap item (49%) than when it applied to the expensive item (32%). This difference was statistically significant ($p=1.01e^{-50}$, d=0.40, 95% CI=[0.35, 0.45]; OR=2.06, 95% CI=[1.87, 2.27]), although the effect size is smaller than the original. However, in our GPT replication with Tversky and Kahneman's original study (N=1,040, with 520 in each condition), all 100% of runs in both conditions answered that they would travel to the far-away store for the discount; see Figure S4. This unexpected uniformity of answers made the statistic we planned to analyse unconstructable, although we can say that our GPT sample's answers did not reveal any evidence of the cognitive bias where decision-making is affected by the base cost of the discounted item rather than just the total discount.

We conducted an unplanned study with the reverse order of answer choices for the question pertaining to the aforementioned scenario (N=1,040, with 520 in each condition). However, the "correct answer" was robust to this order change. In the condition where the cheap item was discounted, all 100% runs still answered that they would travel to the distant store for the discount. In the condition where the expensive item was discounted, an overwhelming 99.8% of surveyed

GPT runs (a total of 1,029) still responded that they would travel to the distant store, whereas only 0.2% of GPT runs (a total of just one) responded that they would remain at the current store and buy the items undiscounted. See Figure S4 for a visual summary.

*STUDY 18.* The paper by Risen & Gilovich (2008) surveyed 120 individuals to examine whether people thought that tempting fate increased the likelihood of bad outcomes, via a survey pertaining to a hypothetical classroom setting. In their original sample, the likelihood of being called upon was seen as higher when the student in question had tempted fate (M=3.43, SD=2.34) than if they had not and had prepared for the class (M=2.53, SD=2.24), $t(116)=2.15$, $p=0.034$, $d=0.39$, 95% CI=[0.03, 0.75]. The Many Labs 2 sample (N=8,000) replicated this effect. The likelihood of being called upon was seen as higher when the student in question had tempted fate (M=4.58, SD=2.44) than if they had prepared for the class (M=4.14, SD=2.45), $t(7998)=8.08$, $p<0.001$, $d=0.18$, 95% CI=[0.14, 0.22]. However, in the total sample (N=1,037) of our study, we find that the likelihood of being called on was judged as higher when the text mentioned having prepared for the class (M=4.75, SD=1.52) compared to when the text mentioned the subject having tempted fate (M=1.99, SD=0.42), $t(1035)=40.0$, $p<0.001$, $d=2.49$, 95% CI=[2.29, 2.68]. This is a very large effect in the opposite direction of both the original finding and the replication.

*STUDY 21.* Hsee (1998) surveyed 83 individuals to provide early evidence for the less-is-better effect. In their original sample, the less expensive scarf gift (M=5.63) was seen as more generous than the more expensive coat gift (M=5.00), $t(82)=3.13$, $p=0.002$, $d=0.69$, 95% CI=[0.24, 1.13], when the cheaper scarf gift was a higher-priced item in its respective category compared to a lower-priced item in the expensive category of coats. The Many Labs 2 sample (N=7,646) replicated this finding; participants in the scarf condition considered their gift more generous (M=5.50, SD=0.89) than did participants in the coat condition, (M=4.61, SD=1.34), t(6569.67)=34.20, p<0.001, d=0.78, 95% CI=[0.74, 0.83]. In our sample (N=1,040), we find that those in the scarf condition saw the gift giver as significantly more generous (M=5.97, SD=0.16) than those in the coat condition (M=3.99, SD=0.26), t(1038)=149, p<0.001, d=9.25, 95% CI=[8.67, 9.82]. This again successfully replicates the original finding, albeit with an extremely large effect size.

**Figure S4**. Subjects' answers to the survey of Tversky and Kahneman (1981) when asked whether—in a situation of needing to buy a cheap item and an expensive item—they would make a trip to a far-away store to buy one of the two at a fixed discount. The "correct answer" of making the trip was robust to reversing the order of answer choices.

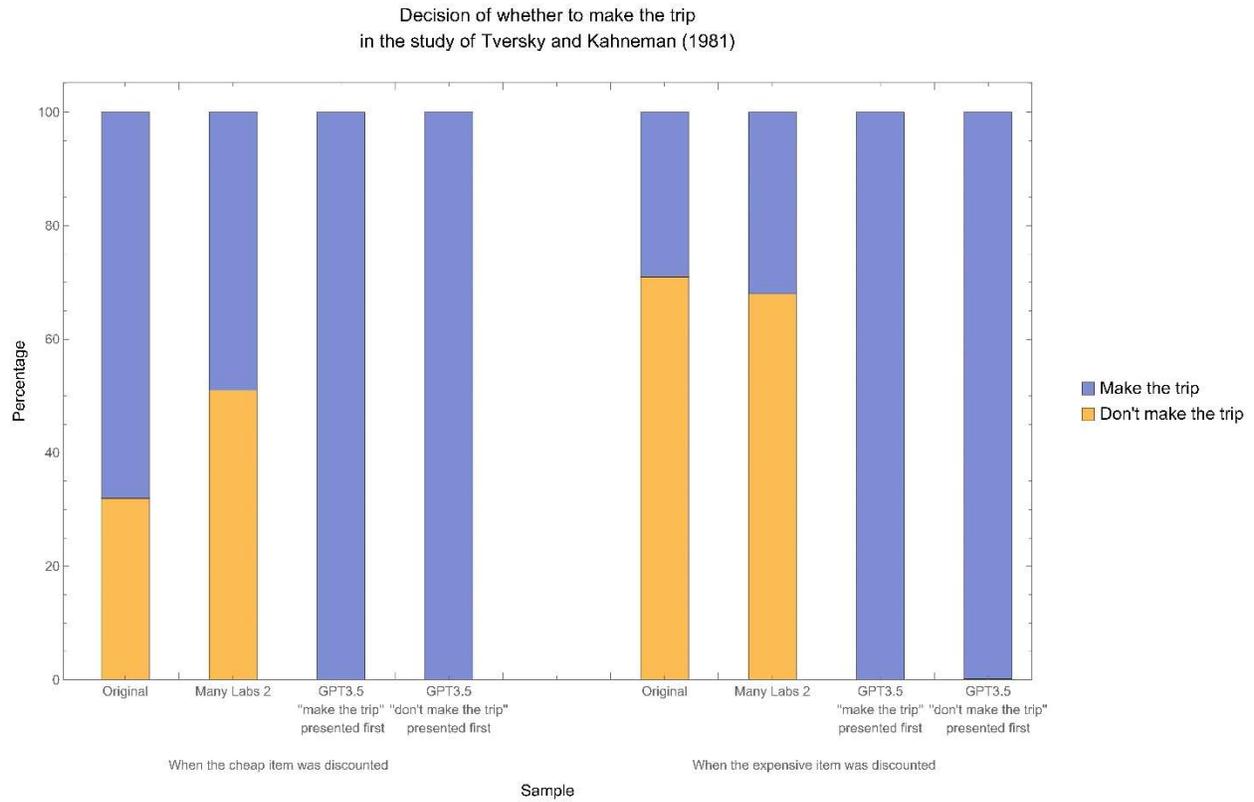

*STUDY 24*. Schwarz et al. (1991) surveyed 100 individuals and found that participants who were asked about their life satisfaction with respect to their relationship specifically before being asked about their life satisfaction in general exhibited a high correlation between their two responses (r=0.67, p<0.05), whereas this correlation was markedly weaker when the two questions were presented in reverse (r=0.32, p<0.05). The difference between the two correlations was statistically significant, z=2.32, p<0.01, q=0.48, 95% CI=[0.07, 0.88]. Many Labs 2 did not replicate this effect, and in fact found the opposite effect. Asking participants about their life satisfaction with respect to their relationship before being asked about their life satisfaction in general resulted in a lower correlation between the two responses (r=0.38) than the other way around (r=0.44). The opposite-direction difference between the correlations was significant, z=-3.03, p=0.002, q=-0.07, 95% CI=[-0.12, -0.02]. In our total sample (N=1,040), we find that the correlation between the two questions was r=-0.446, p=0.001 for when the general question was asked first, and r=-0.490, p<0.001 for when the specific question was asked first. Using the Fisher's r-to-z transformation (Weiss, 2011), we find that this difference was not significant, z=0.906, p=0.365, q=0.056, 95% CI=[-0.066, 0.178]. This is a wholly different pattern of results than both the original study and the Many Labs 2 replication, meaning that we did not replicate the effect in our GPT sample.

*STUDY 25.* Shafir (1994) had 170 participants make decisions between awarding or denying custody to either an average or an extreme parent. In the original sample, participants were more likely to both award (64%) and deny (55%) custody to the parent described with extreme characteristics than they were to the parent described with average characteristics. The sum of these probabilities (119%) was significantly higher than the expected quantity of 100% that one would expect if awarding and denying were complementary, z=2.48, p=0.013, d=0.35, 95% CI=[-0.04, 0.68]. This result indicated that negative features were weighted more heavily than positive ones in the case of rejections and vice versa. The Many Labs 2 sample (N=7,901) did not replicate this effect and in fact provided evidence in favour of an effect in the opposite direction. Participants were less likely to both award (45.5%) and deny (47.6%) custody to the parent described with extreme characteristics, and the sum of these probabilities (93%) was significantly lower than the expected quantity 100%, z=-6.10, p<0.001, d=-0.13, 95% CI=[-0.18, -0.09]. In our sample (N=857), we find that GPT3.5 runs were both more likely to award (95.8%) and deny (100%) custody to the extreme parent than the average parent. Via a logit calculation (Wilson, n.d.), we compute that the sum of these probabilities (195.8%) was significantly higher than the expected quantity of 100%, z=5.11, p<0.001, d=2.11, 95% CI=[1.56, 2.67]. This replicates the original finding that the Many Labs 2 data did not replicate, albeit with a very high effect size.

*STUDY 27.* The original finding by Knobe (2003), which surveyed 78 individuals, was that harmful side effects were judged as more intentional than helpful ones. In the original sample, 82% of participants who were told about an agent (a company's board chairman) whose decision brought about a harmful side effect said the agent did so intentionally, while 23% of participants in the analogous helpful-side-effect condition said that the agent brought about the side effect intentionally, $\chi^2(1, N=78)=27.2$, p<0.001, d=1.45, 95% CI=[0.79, 2.77]. This was replicated by Many Labs 2, which used a seven-point scale of intentionality rather than a yes-no scale. Also, in the Many Labs 2 sample, those in the harmful-side-effect condition (M = 6.03, SD = 1.26) placed a higher degree of blame than the degree of praise (M = 2.54, SD = 1.60) in the helpful-side-effect condition, $t(7553.82) = 108.15$, $p < 1.68e^{-305}$, $d = 2.42$, 95% CI = [2.36, 2.48]. Our sample (N=1,040) consisted of 520 surveyed runs in each condition, using the original version of the survey that used a yes-no scale as well as the seven-point scale for blame/praise. However, the question on rating blame/praise on the seven-point scale saw "correct answers" for both conditions. In the positive-side-effect condition, 99.2% of surveyed GPT runs (a total of 516) described the positive side effect as deserving of a degree seven of praise, or "a lot of praise"; 0.2% of them (a total of just one) described it as deserving of a degree six of praise; and 0.6% of them (a total of 3) described it as deserving of a degree one of praise, or "no praise." In the negative-side-effect condition, 100% of surveyed GPT runs described the negative side effect as deserving of a degree seven of praise, or "a lot of praise." Our pre-registered analysis plan was made unsuitable by the unexpected uniformity of GPT3.5's answers.

We conducted an unplanned and exploratory follow-up study with the reverse order of answer choices for the question pertaining to how much blame/praise is deserved (N=1,040, with 520 in each condition). Specifically, we presented the answer choices in reverse order, from 'A Lot of Praise' to 'No Praise.' In the positive-side-effect condition, the reversal of order resulted in only 3.5% of surveyed GPT runs giving the original "correct answer" of 'A Lot of Praise' (a total of just 18 runs), 0.2% of GPT runs answering with the second-highest level of praise (a total of just one GPT run), and the remaining 96.3% of GPT runs answering with 'No Praise' (a total of 501 runs). The original "correct answer" of 'A Lot of Praise' did not replicate. However, in the negative-side-effect condition, the reversal of order resulted in 93.7% of surveyed GPT runs giving the original "correct answer" of 'A Lot of Praise' (a total of 487 runs), 0.2% of runs responding with the second-highest degree of praise (a total of just one run), and 6.2% of runs responding "No Praise (a total of just 32 runs). The original "correct answer" of 'A Lot of Blame' successfully replicated after reversing the order of answer choices.

For a visual plot of the Cohen's d effect sizes and the corresponding 95% confidence intervals corresponding to the relevant subset of our studies, see Figure S5.

We conducted another exploratory study as a follow-up. The purpose of this second follow-up was to test one of our design trade-offs, which is that we used exclusively temperature as a source of variation in output. As helpfully pointed out by two anonymous reviewers, one reason for the "correct answer" effect may be due to the prompt not including any demographic information, thus artificially restricting variation of the LLM responses compared with a diverse set of human participants. To investigate this concern, we replicated one of the studies that showed a "correct answer" effect to address this concern. Specifically, we re-ran Study 11's (Hauser et al. 2007) condition, where participants are asked to evaluate the moral permissibility of pushing a large man in front of a trolley to save the lives of five others. In this follow-up, we added a randomly selected combination of demographic characteristics to the prompt, instructing the LLM to "respond as a…" The categories that we randomly selected this prompt addition from were gender (male, female), age (20, 30, …, 70), religion (Christianity, Islam, Judaism), ethnicity (Black, White, Asian, Hispanic), and education (High School, College, Advanced Degree). The rest of the prompt remained the same for the condition that we were testing. We collected a total of 982 responses. There were 384 unique combinations of demographic variables (out of a maximum of 432 combinations). Across all combinations, we found strong evidence for the correct answer effect, as 100% of responses indicated that shoving the large man on the tracks to save five others was morally impermissible. This provides evidence that our initial result is unlikely to be primarily explained by the source of variation coming from the temperature setting.

**Figure S5**. Cohen's d effect sizes (with 95% confidence intervals) for the original study, the Many Labs 2 replication, and our GPT3.5 re-replication. The figure excludes the study of Inbar et al. (2009) on whether disgust sensitivity predicts homophobia and the study of Schwarz et al. (1991) on the effects of assimilation and contrast in sequences of questions, because the original studies used Cohen's q instead of Cohen's d. The figure also excludes studies for which the effect size is too large to plot with visual convenience (greater than 3.0). This excluded us from plotting the unprecedentedly high effect size of d=9.25 for our replication of the study of Hsee (1998) on the less-is-better effect, due to the "correct answer" effect of GPT3.5 responding to the inputs of different conditions with much more predeterminedly different answers than did human subjects.

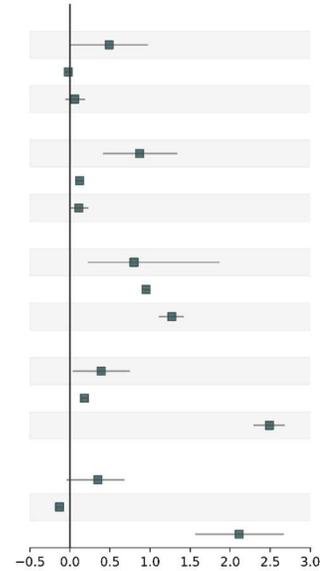